\documentclass[referee]{aa}  

\usepackage{graphicx}
\begin{document}

\title{Na-O Anticorrelation and Horizontal Branches. V. The Na-O anticorrelation in \object{NGC 6441} 
from Giraffe spectra
\thanks{Based on data collected at the European Southern Observatory with
the VLT-UT2, Paranal, Chile (ESO Program 073.D-0211)}}

\author{R.G.~Gratton\inst{1},
        S.~Lucatello\inst{1},
        A.~Bragaglia\inst{2},
        E.~Carretta\inst{2},
	S.~Cassisi\inst{3}
        Y.~Momany\inst{1},
	E.~Pancino\inst{2},
        E.~Valenti\inst{2,4},
	V.~Caloi\inst{5},
	R.~Claudi\inst{1},
	F.~D'Antona\inst{6},
	S.~Desidera\inst{1},
	P. Fran{\c c}ois\inst{7},
	G.~James\inst{7},
	S.~Moehler\inst{8},
	S.~Ortolani\inst{8},
	L.~Pasquini\inst{8},
	G.~Piotto\inst{9},
	A.~Recio-Blanco\inst{10},
}

\offprints{R.G.~Gratton, raffaele.gratton@oapd.inaf.it}

\institute{
	   INAF-Osservatorio Astronomico di Padova, Vicolo dell'Osservatorio 5,35122 Padova, ITALY
\and
           INAF-Osservatorio Astronomico di Bologna, Via Ranzani 1, 40127 Bologna, ITALY
\and
	  INAF-Osservatorio Astronomico di Teramo, via M. Maggini, 64100 Teramo, ITALY
\and
           Dipartimento di Astronomia, Universit\`a di Bologna, Via Ranzani 1, 40127 Bologna, ITALY
\and
       	Istituto di Astrofisica Spaziale e Fisica Cosmica, via Fosso del Cavaliere, 00133 Rome, ITALY
\and
	INAF-Osservatorio Astronomico di Roma, Via Frascati 33, 00040 Monteporzio, ITALY
\and
	European Southern Observatory, 3107 Alonso de Cordova, Vitacura, Casilla 19001, Santiago 19, CHILE
\and
	European Southern Observatory, Karl-Schwarzschild-Strasse 2, D-85748 Garching bei Munchen, GERMANY
\and
	Dipartimento di Astronomia, Universit\`a di Padova, Vicolo dell'Osservatorio 2, 35122, Padova, ITALY	
\and
	Observatoire Astronomique de la C\^ote d'Azur, Boulevard de l'Observatoire, BP 4229, 06304 Nice Cedex 4, FRANCE
}

\date{Received: ; accepted: }

\abstract{}
{We present an analysis of FLAMES-Giraffe spectra for
several bright giants in \object{NGC 6441}, to investigate the
presence and extent of the Na-O anticorrelation in this anomalous
globular cluster.}
{The field of \object{NGC 6441} is very crowded, with severe contamination by
foreground (mainly bulge) field stars. Appropriate membership criteria
were devised to identify a group of 25 likely cluster
members among the about 130 stars observed. Combined with the UVES
data obtained with the same observations (Gratton et
al. 2006), high dispersion abundance analyses are now available for a
total of 30 stars in \object{NGC 6441}, 29 of them having data for both O and
Na. The spectra were analyzed by a standard line analysis procedure;
care was taken to minimize the impact of the differential interstellar
reddening throughout the cluster, and to extract reliable information
from crowded, and moderately high S/N (30-70), moderately high resolution
($R\sim 23,000$) spectra.}
{\object{NGC 6441} has the typical abundance pattern seen in several other
globular clusters. It is very metal-rich ([Fe/H]=$-0.34\pm 0.02\pm
0.04$ dex). 
 There is no clear sign of star-to-star scatter in the Fe-peak
 elements. The $\alpha-$elements Mg, Si, Ca, and Ti are overabundant
 by rather large factors, suggesting that the cluster formed from
 material enriched by massive core collapse SNe. The O-Na
 anticorrelation is well defined, with about 1/4 of the stars being
 Na-rich and O-poor. One of the stars is a Ba-rich
 and moderately C-rich star. Such stars are rare in globular
 clusters.}  {The distribution of [Na/O] ratios among RGB stars in
 \object{NGC 6441} appears similar to the distribution of colors of stars along
 the horizontal branch. The fraction of Na-poor, O-rich stars found in
 \object{NGC 6441} agrees well with that of stars on the red horizontal branch
 of this cluster (in both cases about 80\%), with a sloping
 distribution toward lower values of [O/Na] (among RGB stars) and
 bluer colors (among HB stars).
}

\authorrunning{Gratton R.G. et al.}
\titlerunning{Na-O anticorrelation in NGC6441}

\maketitle

%


\section{INTRODUCTION}


Extensive studies by several groups during the last decades have shown
that globular clusters (GCs) have a peculiar pattern in their chemical
abundances. While they generally are very homogeneous
insofar Fe-peak elements are concerned, they very often (possibly
always) exhibit large star-to-star variations in the abundances of the
light elements (see Gratton et al. 2004). The most prominent feature
is the presence of anticorrelations between the abundances of various
elements: C and N, Na and O, Mg and Al. These anticorrelations are
attributed to the presence at the stellar surfaces of a fraction of
the GC stars of material which has been processed by H burning at
temperatures of a few tens million K. At this temperature, H-burning
occurs through the CNO cycle, so that the abundance pattern of these
elements is shifted toward the equilibrium values, which means
enhanced N and depleted C and O abundances. At the same temperatures,
proton captures on Ne and Mg produce large amounts of Na and Al
(Denissenkov and Denissenkova 1990; Langer et al. 1993), so that the
whole pattern of anticorrelations is present. This pattern is typical
of GC stars; field stars only show changes in C and N abundances
expected from typical evolution of low mass stars (Gratton et
al. 2000; Sweigart \& Mengel 1979; Charbonnel 1994). It is now well
accepted that the abundance pattern seen in GC stars is
primordial, since it is observed in stars at all evolutionary phases
(Gratton et al. 2001, and several other references cited in Gratton et
al. 2004).

Since high Na and low O abundances are the signatures of material
processed through hot H-burning, we expect that this abundance anomaly
be accompanied by high He-contents. D'Antona \& Caloi (2004) estimated
an He excess of $\Delta$Y$\sim 0.04$\ for the Na-rich, O-poor
stars. Values of $\Delta$Y$\sim$0.15 have been recently suggested to justify 
the observed sequences in \object{NGC 2808} (D'Antona et al. 2005). 
While such a difference in the He-content should have small
impact on the colors and magnitudes of stars up to the tip of the red
giant branch (RGB hereafter), a large impact is expected on the colors of the
horizontal branch (HB) stars: He-rich stars should be less
massive by about $0.05~M_\odot$. In the case of GCs of intermediate
metallicity ([Fe/H]$\sim -1.5$), the expectation is then that the
progeny of He-rich, Na-rich, O-poor RGB stars should reside on the
blue part of the HB (i.e. bluer than the RR Lyrae instability strip),
while the progeny of the "normal" He-poor, Na-poor, O-rich stars would
fall within or redward of the instability strip. When comparing
different clusters, the actual pattern may be more complicated,
since small age differences of $\sim 2-3$~Gyr may also cause different
mean colors for the HB stars. However, within a single cluster it is
expected that there should be a correlation between the distribution
of masses (i.e. colors) of the HB-stars and the distribution of Na
and O abundances. Note that
star-to-star variable mass loss is a possibility, possibly fudging the 
correlation.

In this respect, GCs of high metallicity are of great interest. In
the scenario devised by D'Antona \& Caloi
(2001) the He-poor, O-rich, Na-poor stars should lie on the blue side
of the RR Lyrae instability strip; i.e. these clusters should have a
red HB. However, if the cluster age is large, He-rich, O-poor, Na-rich
stars might fall within the instability strip or even be bluer than
that, while in somewhat younger clusters, even He-rich stars would be
on the red HB. Several metal-rich GCs, including the archetypes 47 Tuc
and M71, indeed show short red HBs, even though they also exhibit a
clear O-Na anticorrelation. However, they are probably about 2 Gyrs
younger than the oldest GCs (see Rosenberg 1999; Gratton et al. 2003a;
De Angeli et al. 2005). There are however two other metal-rich GCs
(NGC6388 and \object{NGC 6441}) which show very different HBs: while most of
the stars still lie on the blue side of the HB, both clusters
have a large population of blue HB stars (Rich et
al. 1997). It is very tempting to correlate this feature with the
presence of He-rich stars, that could also explain the fact
that the blue HB is brighter than the red HB (Sweigart \& Catelan
1998), coupled with a rather old age\footnote{A high He content for the blue HB stars of \object{NGC6388} 
is supported by the results of Moehler \& Sweigart (2006)}. Age determinations for these clusters
require analyses of deep color-magnitude diagrams (CMDs hereafter),
 and it is complicated by
the fact that both are projected toward the central regions of our
Galaxy, so that they are severely contaminated by field stars and
affected by large and variable interstellar absorption. On the other
hand, it would be extremely interesting to study the Na-O
anticorrelation in these clusters.

To check if the scenario by D'Antona \& Caloi is acceptable, we
have undertaken an extensive study of the O-Na anticorrelation in
several GCs, with the purpose of determining as accurately as possible
the distribution of stars along this anticorrelation. We expect that
this distribution reflects a distribution in He abundances, hence
 in masses for RGB and HB stars. In this study we exploit the
possibility to obtain high resolution spectra for large number of
stars offered by the FLAMES multifibre facility at VLT-UT2
(Pasquini et al. 2002). With FLAMES, we may simultaneously obtain
spectra of moderately high resolution ($R\sim 23,000$) for more than a
hundred stars using the Giraffe spectrograph, and higher resolution
spectra ($R\sim 45,000$) with larger spectral coverage of up to 8
stars with the UVES spectrograph. This instrument is then ideal for
the present purposes. Early results from this survey, concerning the
intermediate metal-poor clusters \object{NGC 2808} and \object{NGC 6752}, were already
presented by Carretta et al. (2006a and 2006b): the distributions of
O-Na abundances derived for these two clusters closely match
the distributions of stars along the HB. However, more data are
clearly needed before sound conclusions can be drawn. We included in
our survey also \object{NGC 6388} and \object{NGC 6441}. The results from the UVES spectra
of the latter were presented in Gratton et al. (2006, hereinafter
Paper II): \object{NGC 6441} is metal rich
([Fe/H]=$-0.39\pm 0.02\pm 0.04$), and overabundant in the
$\alpha-$elements. We also found that one of the five stars member of
the cluster has Na and O abundances distinctly different (respectively
higher and lower for Na and O) from the remaining four. In the
present paper, we present the analysis of the Giraffe spectra. While
these spectra have lower resolution and cover narrower spectral ranges
than those obtained with UVES, the much larger number of stars
observed gives the opportunity of a better discussion of the
distribution of stars along the O-Na anticorrelation. Unluckily, the
large number of contaminating field stars (which could not be excluded
a priori due to the lack of an appropriate membership study before our
observations were carried out) limited the observed sample to a total
of 25 stars which are bona fide members of \object{NGC 6441} that we combined
with the five bona fide members observed with UVES.
While this is not enough for a detailed comparison
like that performed by Carretta et al. (2006a and 2006b) for the much
easier clusters \object{NGC 2808} and \object{NGC 6752}, it is still enough to give a
first sketch of the distribution.

This paper is organized as follows: in Section 2 we describe the
observational data. In Section 3 we discuss the radial velocities and
the cluster membership. In Section 4 we present the abundance analysis
for the stars found to be members of the cluster, in Section 5 we
discuss our results and we compare the abundance
distributions with other cluster properties. In Section 6 we
comment about one particular star member of \object{NGC 6441}, that 
belongs to the class of Ba-rich stars, quite rare among GCs. Finally,
conclusions are drawn in Section 7.

\section{OBSERVATIONS AND DATA REDUCTION}


\begin{table*}
\begin{center}
\caption{Journal of observations}
\begin{tabular}{lccccc}
\hline
Grating       &    Date    &   Time   & Exp. Time & Seeing   & Airmass \\ 
Configuration &            &          & (sec)     & (arcsec) &         \\
\hline
HR11          & 2004-07-06 & 04:21:42 & 5300 & 1.59 & 1.040-1.174 \\
              & 2004-07-11 & 02:48:19 & 5300 & 1.35 & 1.029-1.052 \\ 
              & 2004-07-11 & 04:19:18 & 5300 & 0.91 & 1.055-1.221 \\ 
HR13          & 2004-07-17 & 05:18:20 & 5300 & 0.69 & 1.202-1.605 \\ 
              & 2004-07-26 & 03:39:54 & 5300 & 1.04 & 1.077-1.282 \\ 
\hline
\end{tabular}
\label{t:journal}
\end{center}
\end{table*}

Observations of \object{NGC 6441} are described in detail in Paper II; we
give here only a few details relevant for the present
purposes. 
Observations were done with
two Giraffe setups, the high-resolution gratings HR11 (centered at
5728~\AA) and HR13 (centered at 6273~\AA) to measure the Na doublets
at 5682-88~\AA~ and 6154-6160~\AA, and the [O{\sc I}] forbidden lines at
6300, 6363~\AA, respectively. Resolution at the center of spectra 
is R=24200 (for HR11) and
R=22500 (for HR13). We have a total exposure time of 15900
second for HR11 and 10600 second for HR13: the latter
were however obtained in better observing conditions, so
that the spectra obtained with HR13 were of higher S/N.

Our targets were selected among isolated RGB stars,
using the photometry described in Paper II. Criteria used to select the
stars are described at length in Paper II. Not all
the stars were observed with both gratings; on a grand total of 127
stars (25 cluster members: see below for the adopted
membership criteria), we have 97 objects (15 cluster members) with
spectra for both gratings, 19 (1 cluster member) with only HR11
observations, and 21 (9 cluster members) with only HR13
observations. Since the Na doublet at 6154-60~\AA\ falls into the
spectral range covered by HR13, we could measure Na abundances for all
target stars, whereas we could expect to measure O abundances only up
to a maximum of 118 stars (24 cluster members).  Table~\ref{t:journal}
lists information about the two pointings.

\begin{figure}[h]
\includegraphics[width=8.8cm]{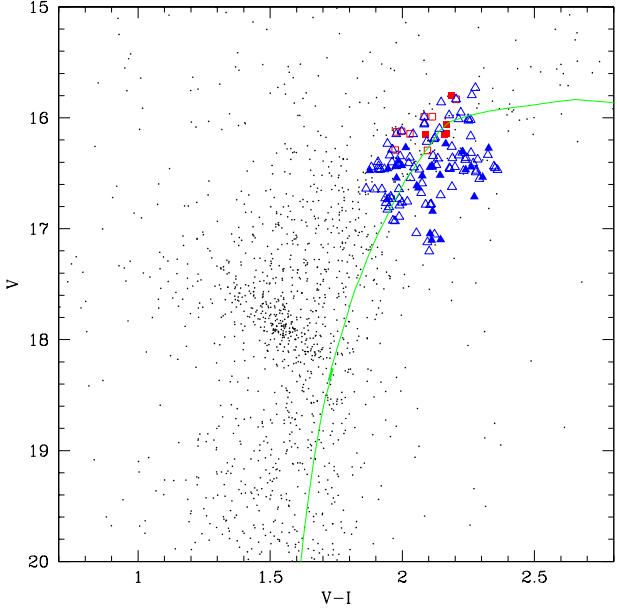}
\caption[]{$(V,V-I)$ color magnitude diagram for selected stars in the field of \object{\object{NGC 6441}} 
(from Valenti et al. 2006). Squares indicate stars observed with FLAMES--UVES, while triangles are 
the stars targeted by FLAMES--Giraffe. Filled symbols 
mark stars member of the cluster on the basis of RVs and location in 
the field close to the cluster center; open symbols are non-member stars.
Overimposed is an isochrone computed for an age of 13\,Gyr and [Fe/H]=-0.32 from Pietrinferni 
et al. (2004). This isochrone is for a solar scaled composition.}
\label{f:cmd}
\end{figure}

Table \ref{t:phot} gives details on the main parameters for the member
stars (see next Section for the adopted membership criteria). Star
designations are according to the photometry described in Paper II,
from which photometric data were also taken. Coordinates (at J2000
equinox) are from our astrometry; distances from the cluster center
were obtained considering the nominal position given by Harris
(1996). The signal-to-noise ratios S/N were estimated from the
pixel-to-pixel scatter in spectral regions relatively free from absorption
lines;however, since spectral regions truly free from absorption features are 
virtually nonexistent in these spectra, we adopted this method
after dividing each spectrum by an average
spectrum obtained summing the spectra of all stars members of the
cluster. The $V$, $V-I$. CMD of our sample is
shown in Figure~\ref{f:cmd} with overimposed an appropriate isochrone (13\,Gyr, [Fe/H]=-0.32) 
from Pietrinferni et al. (2004), shifted by the distance modulus 
(m-M)$_V$=16.79 (Harris catalog) and reddened adopting E($B-V$)=0.49. Our targets are in the range from about $V$=16.2 to
17.2 and $V-I$=1.88 to 2.32. The selected stars are well below the tip
of the RGB and span the whole range in color of the broad RGB of \object{NGC 6441}.

We used the 1-d, wavelength calibrated spectra as reduced by the
dedicated Giraffe pipeline (BLDRS v0.5.3, written at the Geneva
Observatory, see {\it http://girlbirds.sourceforge.net}).
The radial velocities (RVs) have been 
measured by the Giraffe pipeline, which performs 
a cross-correlation using an appropriate synthetic spectrum as a template. 
The typical errors on these measurements are around 0.3-0.5\, km/s.
Further analysis was done with IRAF\footnote{IRAF is distributed by the National
Optical Astronomical Observatory, which is operated by the Association
of Universities for Research in Astronomy, under contract with the
National Science Foundation}. We subtracted the background using 8
fibers dedicated to the sky, rectified the spectra, corrected for
contamination by telluric features in the HR13 spectra using the task
TELLURIC and shifted all the spectra to zero RV before summing all 
individual spectra for each stars.  

\section{CLUSTER MEMBERSHIP}

\begin{figure}[h]
\includegraphics[width=8.8cm]{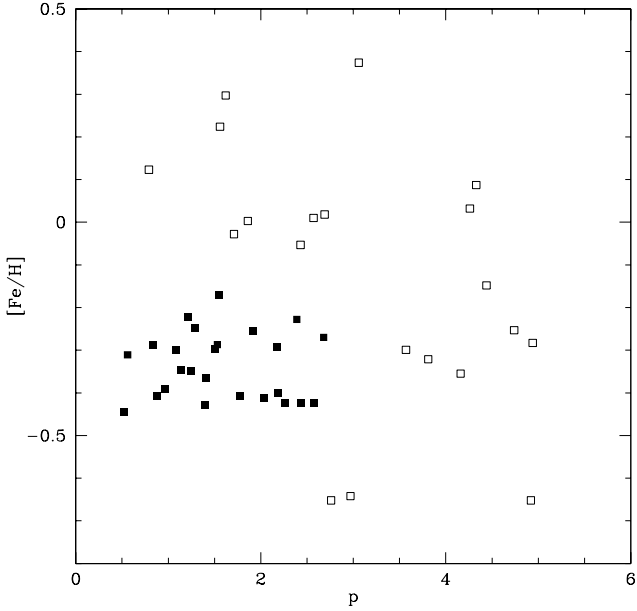}
\caption[]{Derived iron abundance as a function of the p parameter (see text for definition) for the 
analyzed stars. Filled symbols are likely members, open symbols are non members.}
\label{f:pabb}
\end{figure}


\begin{table*}
\begin{center}
\caption{Photometry and spectroscopic data for stars observed with Giraffe. }
\begin{tabular}{lrrrrrrrrrrrr}
\hline
Star & Gratings &  RA     &  Dec.  &  Dist. & V  & V-I  & V-K  & RV & S/N& S/N&$\sigma$(EW)& [Fe/H] \\
     &          &(degree) &(degree)&(arcsec)&(mag)&(mag)&(mag)&(km s$^{-1})$& HR11 & HR13& (m\AA)~~&   \\
\hline
6005198 & 11+13 & 267.3757 & $-$37.0886 & 528 & 16.274 & 2.327 & 5.007 & 12.1 & 15 & 43 & 17.1 & $-$0.41 \\
6005934 & 11+13 & 267.3842 & $-$37.2010 & 727 & 16.387 & 1.985 & 4.359 & 23.0 & 22 & 37 & 13.7 & $-$0.42 \\
6007741 & 11+13 & 267.4520 & $-$37.0056 & 334 & 16.612 & 2.055 & 4.467 & 17.0 &    & 28 & 15.6 & $-$0.25 \\
6010149 & 11    & 267.4408 & $-$37.1361 & 446 & 16.842 & 2.114 & 4.510 & 30.6 & 23 &    & 11.0 & $-$0.17 \\
6012768 & 13    & 267.4320 & $-$37.1921 & 616 & 17.043 & 2.105 &       & 13.3 &    & 50 &  7.8 & $-$0.42 \\
7004955 & 13    & 267.5810 & $-$37.0124 & 160 & 16.232 & 2.164 & 4.783 & 11.4 &    & 69 & 10.0 & $-$0.22 \\
7006255 & 11+13 & 267.5414 & $-$37.0654 &  62 & 16.430 & 1.998 &       & 13.3 & 33 & 50 &  8.3 & $-$0.39 \\
7006305 & 11+13 & 267.6267 & $-$37.0145 & 248 & 16.437 & 2.112 & 4.576 & 27.0 & 34 & 65 &  9.8 & $-$0.29 \\
7006319 & 11+13 & 267.5162 & $-$37.0608 & 113 & 16.438 & 1.975 & 4.483 &  6.5 & 45 & 58 & 11.4 & $-$0.29 \\
7006354 & 13    & 267.5631 & $-$37.1049 & 196 & 16.442 & 2.261 & 4.842 & 32.4 &    & 53 &  9.8 & $-$0.41 \\
7006377 & 11+13 & 267.5129 & $-$37.1338 & 320 & 16.445 & 2.105 & 4.646 & 13.5 & 41 & 55 &  6.5 & $-$0.37 \\
7006470 & 11+13 & 267.5570 & $-$36.9669 & 303 & 16.458 & 1.946 & 4.423 &  0.9 & 31 & 55 &  9.9 & $-$0.29 \\
7006590 & 11+13 & 267.4963 & $-$37.0236 & 192 & 16.475 & 1.877 &       &  4.3 & 30 & 50 & 10.1 & $-$0.41 \\
7006935 & 11+13 & 267.5364 & $-$36.9426 & 394 & 16.518 & 2.143 & 4.687 & 18.9 & 42 & 57 &  8.0 & $-$0.43 \\
7006983 & 11+13 & 267.6414 & $-$36.9409 & 470 & 16.522 & 2.075 & 4.441 & 27.8 & 42 & 57 &  6.8 & $-$0.23 \\
7007064 & 11+13 & 267.5956 & $-$37.0473 & 122 & 16.535 & 2.304 & 4.953 & 29.1 & 49 & 66 & 10.1 & $-$0.45 \\
7007118 & 13    & 267.5242 & $-$36.9683 & 310 & 16.542 & 1.978 & 4.367 & 18.9 &    & 64 &  5.7 & $-$0.35 \\
7007884 & 13    & 267.5766 & $-$37.0884 & 150 & 16.631 & 2.061 & 4.955 & 28.2 &    & 64 & 11.6 & $-$0.31 \\
7008891 & 13    & 267.5674 & $-$37.0043 & 173 & 16.730 & 1.989 & 4.312 & 13.6 &    & 54 &  7.8 & $-$0.30 \\
7013582 & 13    & 267.6406 & $-$36.9898 & 334 & 17.097 & 2.145 & 4.586 & 17.8 &    & 53 &  7.4 & $-$0.35 \\
8005158 & 13    & 267.7439 & $-$36.9866 & 594 & 16.267 & 2.013 & 4.480 & 36.8 &    & 51 &  8.3 & $-$0.40 \\
8005404 & 11+13 & 267.6974 & $-$37.1418 & 527 & 16.306 & 2.228 & 4.894 & 48.4 & 26 & 59 & 12.3 & $-$0.42 \\
8006535 & 11+13 & 267.6661 & $-$36.9818 & 408 & 16.467 & 2.240 & 4.973 &  8.1 & 33 & 88 & 10.8 & $-$0.25 \\
8008693 & 11+13 & 267.6896 & $-$36.9899 & 449 & 16.711 & 2.272 & 4.850 & 25.0 & 34 & 32 & 25.8 & $-$0.30 \\
8013657 & 13    & 267.7280 & $-$37.1704 & 660 & 17.103 & 2.111 & 4.632 & 44.0 &    & 41 &  6.8 & $-$0.27 \\
\hline
\end{tabular}
\label{t:phot}
\end{center}
\end{table*}

\begin{figure}[h]
\includegraphics[width=8.8cm]{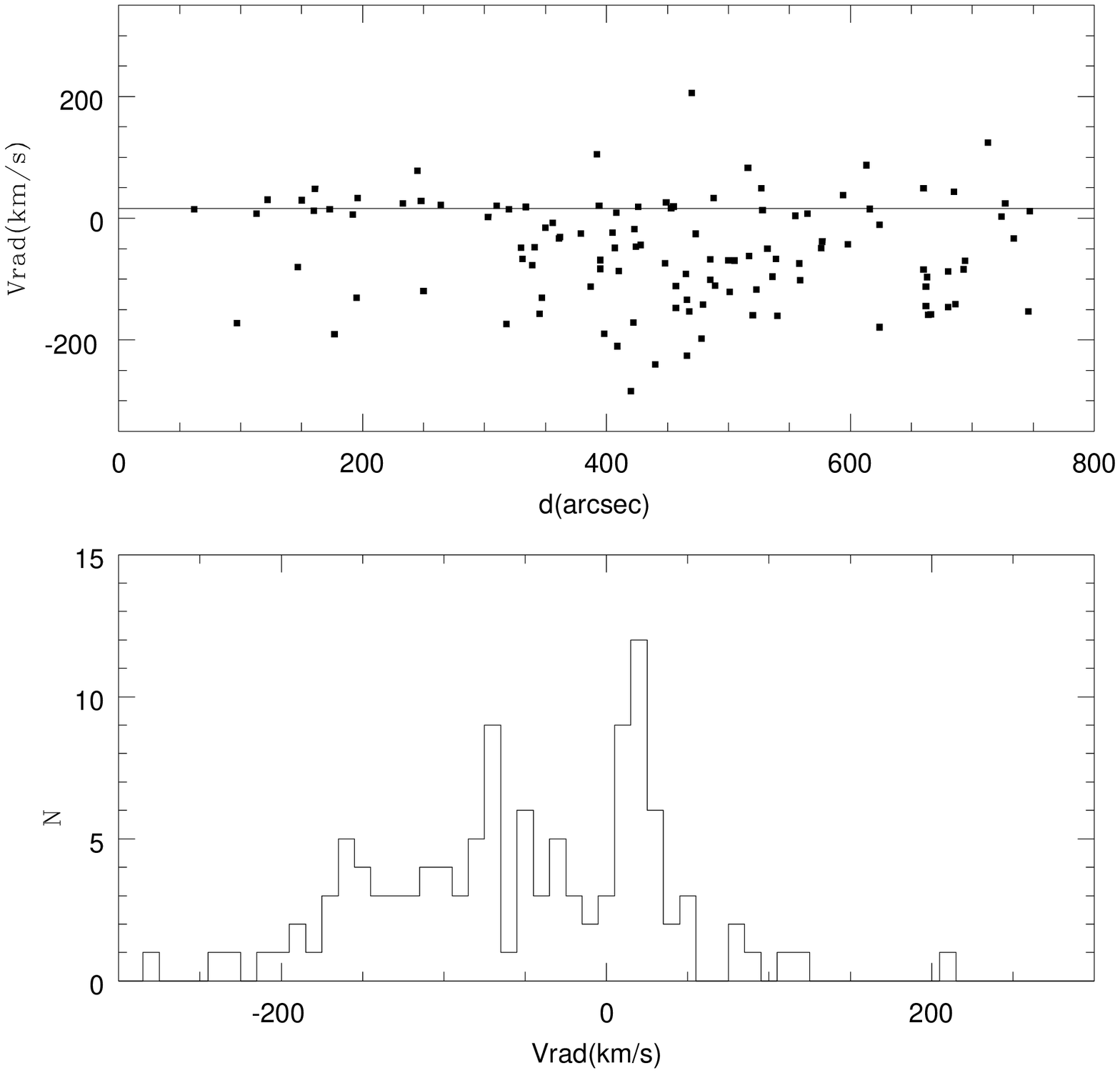}
\caption[]{Upper panel: RV vs distance from the cluster center (in arcsec) for all the observed 
stars. Lower panel:
radial velocity histogram for all the observed stars. The excess of stars at
$RV\sim +20$~km~s$^{-1}$\ corresponds to the average velocity of \object{NGC 6441}; this is
represented as a solid line in the upper panel.}
\label{f:istvel}
\end{figure}

In order to properly define membership to the cluster, we first
defined a parameter $p=\sqrt{(Dist/300)^2+((RV-21.0)/12.5)^2}$, where
$Dist$\ is the projected distance from the cluster center (in arcsec),
and $RV$\ is the radial velocity in km~s$^{-1}$, $p$\ represents
the distance of each star from the cluster center mean
position in the $Dist-RV$\ plane, roughly expressed in units of
standard deviation of the distributions. Note that according to Harris
(1996) the tidal radius of \object{NGC 6441} is about 467 arcsec, and the
average radial velocity is $+16.4\pm 1.2$~km~s$^{-1}$; however the
final values we adopted for the average radial velocity and its
scatter were iteratively obtained from the average and r.m.s. scatter
of the probable members identified in our sample. $p$\ is obviously
related to the probability that a star is a member of the
cluster. Assuming uniform distributions in both position and RVs
 (for this last, over the range $\pm 190$~km~s$^{-1}$\ from
the mean cluster velocity, corresponding to the observed spread in
RVs), we would expect 0.46 field stars with $p<1$, 3.24
with $1<p<2$, 6.39 with $2<p<3$, and 3.70 with $3<p<4$\footnote{This
value is smaller than for the preceding bin because the edge of the
Oz-Poz field was reached at this distance}, while the observed numbers
in these bins are 8, 18, 14, and 3 (where we included also stars
observed with UVES): the excess of objects around the cluster position
in the $Dist-RV$\ plane is obvious. These numbers suggest that
membership probability based on these criteria alone is $\sim 94$\%
for $p<1$, $\sim 82$\% for $1<p<2$, about a half for $2<p<3$, and
smaller for larger values of $p$. Combining this datum with
metallicity we may improve separation of likely members from field
stars, of course with some risk of biasing the metallicity
distribution. We then examined the chemical composition for all stars
with $p<5$: the results are graphically shown in Figure~\ref{f:pabb}. After
inspection of this Figure, we finally defined as likely members of
\object{NGC 6441} those stars with $p<3.0$\ and $-0.6<$[Fe/H]$<-0.1$. 25 stars
passed this criterion. Note that we removed from the sample 1 star
out of 8 with $p<1$, 4 stars out of 18 with $1<p<2$, and 5 stars out of 14
with $2<p<3$, in quite good agreement with the expected number of
field stars. All stars with $p<2.5$\ removed from the sample of
likely members are much more metal rich than the cluster average;
the field stars have a mean metallicity around
[Fe/H]$\sim -0.1$\footnote{As noticed in Paper II, most of the field
stars should belong to the bulge. The line of sight toward \object{NGC 6441}
passes at about 1.1 kpc from the cluster center. However, the field
star average metallicity might be biased by the selection procedure
adopted, and should not be taken as representative of the bulge metal
abundance in the direction toward \object{NGC 6441}.}. There might be some
additional members of \object{NGC 6441}: this is
suggested by the height of the peak in the radial velocity
distribution around the cluster mean velocity (Figure~\ref{f:istvel}).
 However, their membership being doubtful, we prefer not to
use them in our analysis. Note that all stars considered members or
probable members of \object{NGC 6441} in Paper II would be considered
members, and all those considered as field stars would be field
stars according to the criteria of the present paper\footnote{The list
of stars not members of \object{NGC 6441} can be obtained upon request from the
authors.}.

The average radial velocity that we obtained from our 25 confirmed
members is $21.0\pm 2.5$~km~s$^{-1}$, with a star-to-star
r.m.s. scatter of 12.5~km~s$^{-1}$. If we add the five stars members of
the cluster observed in Paper II, the average radial velocity is the
same, and the r.m.s. scatter increases to 13.2~km~s$^{-1}$.  
The observed radial velocity scatter, although quite large, is not surprising 
for this cluster, which is known to have a very large central velocity 
dispersion (see e.g. Illingworth 1979, Dubath et al. 1997) 
 The observed mean radial velocity value
is slightly larger than that given by Harris (1996); this might be due
to some contamination by the field stars in some of the literature
determinations used by Harris, since on average field stars have a
negative radial velocity of $\sim -50$~km~s$^{-1}$\ (see Figure~\ref{f:istvel}), in agreement 
with the expected values for this direction on the basis of the bulge rotation
curve by Tiede \& Tendrup (1997) (see also C{\^o}t{\'e} 1999).

\begin{table*}
\caption{Equivalent Widths from Giraffe spectra (in electronic form)}
\tiny
\begin{tabular}{lrrrrrrrrrrrrr}
\hline
El   & Wavel.  & E.P. & log gf &6005198&6005934&6007741&6010149&6012768&7004955&7006255&7006305&7006319&7006354\\
     & (\AA)   & (eV) &       &(m\AA)~~&(m\AA)~~&(m\AA)~~&(m\AA)~~&(m\AA)~~&(m\AA)~~&(m\AA)~~&(m\AA)~~&(m\AA)~~&(m\AA)~\\
\hline
 8.1 & 6300.31 & 0.00 &  -9.75 &  45.0 &  54.0 &   0.0 &   0.0 &   0.0 &  36.5 &   0.0 &  57.4 &  44.0 &  56.0 \\
 8.1 & 6363.79 & 0.02 & -10.25 &  38.0 &  28.0 &  14.0 &   0.0 &  16.0 &   0.0 &  11.0 &  18.0 &  29.0 &  36.0 \\
11.1 & 6154.23 & 2.10 &  -1.57 & 126.7 &  78.4 & 117.8 &   0.0 &  83.4 & 163.3 & 116.3 & 122.3 &  74.6 & 114.7 \\
11.1 & 6160.75 & 2.10 &  -1.26 & 141.0 & 103.7 & 133.4 &   0.0 &   0.0 & 182.8 & 147.2 & 139.4 & 102.6 & 126.2 \\
12.1 & 5711.09 & 4.34 &  -1.71 &   0.0 & 124.8 &   0.0 & 154.0 &   0.0 &   0.0 & 113.5 & 158.6 & 143.0 &   0.0 \\
12.1 & 6318.71 & 5.11 &  -1.97 &  63.8 &  72.5 & 100.3 &   0.0 &  52.0 &  74.3 &  75.6 &  84.6 &  85.3 &  70.5 \\
12.1 & 6319.24 & 5.11 &  -2.20 &  57.5 &  66.4 &  54.0 &   0.0 &  47.7 &  68.1 &  63.0 &  74.0 &  80.7 &  63.7 \\
14.1 & 5645.62 & 4.93 &  -2.14 &  37.7 &  57.8 &   0.0 &  84.6 &   0.0 &   0.0 &  53.2 &  65.6 &  64.9 &   0.0 \\
14.1 & 5665.56 & 4.92 &  -2.04 &  90.0 &  96.1 &   0.0 &  97.1 &   0.0 &   0.0 &  58.6 &  49.3 &  67.3 &   0.0 \\
14.1 & 5684.49 & 4.95 &  -1.65 & 109.2 &  57.2 &   0.0 &  75.2 &   0.0 &   0.0 &  75.8 &  69.6 &  77.4 &   0.0 \\
14.1 & 5690.43 & 4.93 &  -1.87 &  38.9 &  65.5 &   0.0 &  70.8 &   0.0 &   0.0 &  42.2 &  47.1 &  61.7 &   0.0 \\
14.1 & 5793.08 & 4.93 &  -2.06 &  39.9 &  55.3 &   0.0 &  62.2 &   0.0 &   0.0 &  43.9 &  46.7 &  73.3 &   0.0 \\
14.1 & 6125.03 & 5.61 &  -1.57 &  44.3 &  56.5 & 133.6 &   0.0 &  38.3 &  57.9 &  35.5 &  53.3 &  36.1 &  47.6 \\
14.1 & 6145.02 & 5.61 &  -1.44 &  45.3 &  49.5 &  57.2 &   0.0 &  39.9 &  41.4 &  39.2 &  51.9 &  34.3 &  43.0 \\
20.1 & 6161.30 & 2.52 &  -1.27 & 141.9 & 157.2 & 200.2 &   0.0 & 114.4 & 186.4 & 144.0 & 158.3 & 147.8 & 155.6 \\
20.1 & 6166.44 & 2.52 &  -1.14 & 128.3 & 144.1 & 163.0 &   0.0 &  98.3 & 175.8 & 130.3 & 147.7 & 128.3 & 140.7 \\
20.1 & 6169.04 & 2.52 &  -0.80 & 158.8 & 161.9 & 179.6 &   0.0 & 124.9 & 203.7 & 168.5 & 169.6 & 159.0 & 173.3 \\
20.1 & 6169.56 & 2.52 &  -0.48 & 176.2 & 160.5 & 212.7 &   0.0 & 145.4 & 220.9 &   0.0 & 177.6 & 184.3 & 163.2 \\
21.2 & 5640.99 & 1.50 &  -0.86 &  77.2 &  89.9 &   0.0 &   0.0 &   0.0 &   0.0 &  99.2 &  82.4 &  97.9 &   0.0 \\
21.2 & 5667.15 & 1.50 &  -1.11 &  90.4 &  88.9 &   0.0 &   0.0 &   0.0 &   0.0 &  95.4 &  71.2 &  95.4 &   0.0 \\
21.2 & 5669.04 & 1.50 &  -1.00 &  93.4 &  75.8 &   0.0 &   0.0 &   0.0 &   0.0 &  75.0 &  60.4 &  88.5 &   0.0 \\
21.2 & 5671.83 & 1.45 &   0.56 &  71.0 & 119.9 &   0.0 & 149.4 &   0.0 &   0.0 &  97.7 & 100.7 & 115.8 &   0.0 \\
21.2 & 6245.62 & 1.51 &  -1.05 &  90.6 &  92.1 & 104.3 &   0.0 &  59.2 & 116.4 & 102.5 &  88.7 & 100.1 &  86.8 \\
21.2 & 6279.74 & 1.50 &  -1.16 &  82.0 &  89.6 & 114.7 &   0.0 &  59.0 &  94.7 &  94.0 &  75.4 & 100.6 &  63.0 \\
22.1 & 6126.22 & 1.07 &  -1.42 & 178.0 & 136.2 & 216.3 &   0.0 & 113.9 & 201.7 & 150.0 & 156.9 & 153.2 & 178.4 \\
23.1 & 5624.89 & 1.05 &  -1.07 &  62.8 &  75.5 &   0.0 & 108.5 &   0.0 &   0.0 &  85.2 &  66.9 & 104.7 &   0.0 \\
23.1 & 5626.01 & 1.04 &  -1.25 &  88.6 &  95.3 &   0.0 & 157.0 &   0.0 &   0.0 & 112.8 &  96.8 & 112.2 &   0.0 \\
23.1 & 5627.64 & 1.08 &  -0.37 &  76.6 &   0.0 &   0.0 & 155.4 &   0.0 &   0.0 & 110.6 & 102.9 & 131.1 &   0.0 \\
23.1 & 5657.44 & 1.06 &  -1.02 &  64.0 & 100.7 &   0.0 & 117.3 &   0.0 &   0.0 &  86.0 &  89.0 &  92.0 &   0.0 \\
23.1 & 5668.36 & 1.08 &  -1.02 &  74.5 &  92.8 &   0.0 &  89.0 &   0.0 &   0.0 & 104.0 &  84.2 & 101.4 &   0.0 \\
23.1 & 5670.86 & 1.08 &  -0.42 &  91.4 & 136.0 &   0.0 & 128.0 &   0.0 &   0.0 & 128.2 & 112.8 & 133.5 &   0.0 \\
23.1 & 5698.53 & 1.06 &  -0.11 & 124.2 & 109.1 &   0.0 & 192.1 &   0.0 &   0.0 & 126.6 & 117.9 & 172.4 &   0.0 \\
23.1 & 5727.06 & 1.08 &  -0.01 & 108.9 &  98.3 &   0.0 & 192.5 &   0.0 &   0.0 & 130.5 & 165.8 & 179.8 &   0.0 \\
23.1 & 5737.06 & 1.06 &  -0.74 & 107.9 &  91.3 &   0.0 & 154.7 &   0.0 &   0.0 & 119.5 & 130.0 & 129.7 &   0.0 \\
23.1 & 5743.45 & 1.08 &  -0.97 &  78.5 &  75.1 &   0.0 & 150.8 &   0.0 &   0.0 &  92.9 & 124.6 & 132.9 &   0.0 \\
23.1 & 6251.82 & 0.29 &  -1.34 & 216.7 & 135.4 & 178.5 &   0.0 & 113.3 & 218.5 & 194.7 & 193.7 & 197.5 & 212.7 \\
24.1 & 5628.65 & 3.42 &  -0.77 &  31.7 &  83.4 &   0.0 &  63.2 &   0.0 &   0.0 &  49.7 &  43.9 &  61.1 &   0.0 \\
24.1 & 5783.07 & 3.32 &  -0.40 &  75.0 & 105.5 &   0.0 &  72.4 &   0.0 &   0.0 &  70.6 &  57.9 &  74.7 &   0.0 \\
24.1 & 5787.93 & 3.32 &  -0.08 &  79.9 & 101.9 &   0.0 & 105.8 &   0.0 &   0.0 &  90.0 &  89.8 & 107.6 &   0.0 \\
24.1 & 6330.10 & 0.94 &  -2.87 & 183.6 & 117.5 & 154.1 &   0.0 &   0.0 & 166.7 & 161.3 & 163.0 & 162.2 & 175.8 \\
26.1 & 5619.61 & 4.39 &  -1.49 &  48.9 &  78.5 &   0.0 &  71.4 &   0.0 &   0.0 &  38.9 &  53.4 &  68.0 &   0.0 \\
26.1 & 5635.83 & 4.26 &  -1.59 &  58.1 &  63.3 &   0.0 &  91.5 &   0.0 &   0.0 &  68.3 &  62.0 &  72.9 &   0.0 \\
26.1 & 5636.70 & 3.64 &  -2.53 &  48.2 &  64.1 &   0.0 &  70.2 &   0.0 &   0.0 &  71.4 &  62.8 &  68.8 &   0.0 \\
26.1 & 5650.00 & 5.10 &  -0.80 &  38.3 &  63.8 &   0.0 &  65.4 &   0.0 &   0.0 &  54.8 &  62.3 &  73.7 &   0.0 \\
26.1 & 5651.48 & 4.47 &  -1.79 &  38.8 &  46.4 &   0.0 &  53.6 &   0.0 &   0.0 &  46.0 &  58.9 &  51.0 &   0.0 \\
26.1 & 5652.33 & 4.26 &  -1.77 &  46.4 &  62.9 &   0.0 &  55.6 &   0.0 &   0.0 &  52.5 &  39.6 &  63.4 &   0.0 \\
26.1 & 5717.84 & 4.28 &  -0.98 &  71.3 &  76.0 &   0.0 & 121.6 &   0.0 &   0.0 &  85.0 &  99.0 &  87.2 &   0.0 \\
26.1 & 5731.77 & 4.26 &  -1.10 &   0.0 &  62.7 &   0.0 & 104.3 &   0.0 &   0.0 &  85.2 &  94.4 &  88.5 &   0.0 \\
26.1 & 5738.24 & 4.22 &  -2.24 &  38.5 &  66.3 &   0.0 &  61.4 &   0.0 &   0.0 &  41.4 &  54.6 &  41.5 &   0.0 \\
26.1 & 5741.86 & 4.26 &  -1.69 &  66.7 &  50.0 &   0.0 &  87.4 &   0.0 &   0.0 &  53.4 &  59.2 &  67.0 &   0.0 \\
26.1 & 5752.04 & 4.55 &  -0.92 &  74.8 &  54.3 &   0.0 &  87.0 &   0.0 &   0.0 &  85.2 &  74.6 &  65.5 &   0.0 \\
26.1 & 5775.09 & 4.22 &  -1.11 &  64.0 &  79.1 &   0.0 &  92.9 &   0.0 &   0.0 &  73.1 &  75.3 & 102.1 &   0.0 \\
26.1 & 5793.92 & 4.22 &  -1.62 &  69.9 &  48.3 &   0.0 &  77.7 &   0.0 &   0.0 &  64.9 &  51.2 &  78.8 &   0.0 \\
26.1 & 5806.73 & 4.61 &  -0.93 &  49.3 &  83.3 &   0.0 &  89.6 &   0.0 &   0.0 &  77.7 &  91.4 &  88.7 &   0.0 \\
26.1 & 5827.87 & 3.28 &  -3.16 &  56.6 &  57.2 &   0.0 &   0.0 &   0.0 &   0.0 &  63.4 &   0.0 &  74.9 &   0.0 \\
26.1 & 6151.62 & 2.18 &  -3.26 & 137.4 & 125.0 & 153.8 &   0.0 &  96.9 & 166.3 & 123.8 & 140.8 & 113.2 &  91.8 \\
26.1 & 6159.38 & 4.61 &  -1.88 &  57.2 &  32.4 &  65.3 &   0.0 &  85.3 &  38.4 &  33.9 &  48.6 &  24.6 &  51.8 \\
26.1 & 6165.36 & 4.14 &  -1.48 &  70.1 &  93.3 & 105.3 &   0.0 &  76.0 &  96.4 &  66.8 &  86.3 &  58.9 &  83.3 \\
26.1 & 6173.34 & 2.22 &  -2.84 & 145.8 & 154.5 & 174.2 &   0.0 & 119.9 & 190.7 & 152.3 & 147.1 & 146.2 & 154.8 \\
26.1 & 6187.40 & 2.83 &  -4.12 &  47.1 &  42.7 &  50.2 &   0.0 &  33.7 &  57.9 &  40.8 &  49.4 &  33.3 &  42.2 \\
26.1 & 6188.00 & 3.94 &  -1.60 &  78.4 &  84.2 & 103.4 &   0.0 &  74.8 & 108.2 &  81.8 &  93.4 &  68.3 &  91.3 \\
26.1 & 6200.32 & 2.61 &  -2.39 & 125.6 & 138.8 & 167.8 &   0.0 &  62.5 & 171.4 & 139.0 & 137.1 & 131.1 & 141.6 \\
26.1 & 6213.44 & 2.22 &  -2.54 & 158.2 & 165.7 & 183.8 &   0.0 & 134.8 & 200.8 & 171.2 & 166.5 & 174.4 & 173.8 \\
26.1 & 6219.29 & 2.20 &  -2.39 & 181.0 & 156.9 & 171.3 &   0.0 & 125.7 & 196.6 & 180.6 & 187.5 & 178.3 & 186.0 \\
26.1 & 6232.65 & 3.65 &  -1.21 & 117.7 & 131.0 & 170.5 &   0.0 & 129.6 & 155.2 & 127.7 & 134.8 & 136.5 & 134.0 \\
26.1 & 6240.65 & 2.22 &  -3.23 & 139.1 & 119.0 & 138.5 &   0.0 &  94.7 & 142.9 & 135.0 & 121.8 & 103.4 & 121.8 \\
26.1 & 6265.14 & 2.18 &  -2.51 & 202.6 & 155.0 & 185.1 &   0.0 & 150.0 & 210.9 & 209.9 & 198.5 & 193.4 & 204.6 \\
26.1 & 6270.23 & 2.86 &  -2.55 & 101.7 & 116.5 & 129.7 &   0.0 &  78.2 & 133.3 & 136.0 & 103.9 & 123.3 & 109.6 \\
26.1 & 6311.50 & 2.83 &  -3.16 & 101.2 &  74.2 &  96.1 &   0.0 &  70.6 & 117.6 & 105.1 &  99.5 & 101.6 &  95.4 \\
26.1 & 6315.81 & 4.07 &  -1.67 &  62.9 &  73.4 &  80.4 &   0.0 &  54.1 &  74.7 &  74.0 &  67.4 &  75.7 &  70.4 \\
26.1 & 6322.69 & 2.59 &  -2.38 & 144.8 & 122.0 & 129.3 &   0.0 & 107.9 & 152.4 & 162.5 & 140.1 & 145.1 & 156.4 \\
26.1 & 6330.85 & 4.73 &  -1.22 &  46.9 &  60.9 &  53.9 &   0.0 &  37.6 &  52.2 &  42.5 &  63.7 &  67.6 &  56.5 \\
26.1 & 6335.34 & 2.20 &  -2.28 & 216.7 & 152.5 & 183.4 &   0.0 & 158.5 & 218.5 & 211.1 & 215.7 & 207.9 & 227.0 \\
26.1 & 6353.84 & 0.91 &  -6.41 &  71.6 &  46.0 &  81.9 &   0.0 &  41.2 &  77.2 &  69.3 &  61.9 &  64.8 &  60.6 \\
26.1 & 6380.75 & 4.19 &  -1.34 & 102.5 &  68.9 &  89.6 &   0.0 &  73.4 &  86.8 &  85.0 &  91.5 &  96.3 &  93.0 \\
26.1 & 6392.54 & 2.28 &  -3.97 &  97.1 &  52.5 &  77.2 &   0.0 &  60.7 &  63.4 &  96.0 &  81.2 &  85.6 &  88.4 \\
26.2 & 6247.56 & 3.87 &  -2.32 &  23.4 &  39.4 &  26.2 &   0.0 &  38.4 &  26.1 &  23.8 &  33.1 &   0.0 &  38.7 \\
26.2 & 6369.46 & 2.89 &  -4.21 &  26.9 &  24.2 &   0.0 &   0.0 &  53.4 &   0.0 &   0.0 &  31.9 &  30.0 &  28.1 \\
28.1 & 5643.09 & 4.16 &  -1.25 &  31.8 &  29.3 &   0.0 &  54.0 &   0.0 &   0.0 &  23.6 &  47.5 &  45.5 &   0.0 \\
28.1 & 5709.56 & 1.68 &  -2.14 & 107.2 & 160.4 &   0.0 & 178.9 &   0.0 &   0.0 & 160.8 & 130.0 & 177.0 &   0.0 \\
28.1 & 5760.84 & 4.10 &  -0.81 & 114.3 &  60.2 &   0.0 & 100.1 &   0.0 &   0.0 &  85.0 &  82.1 &  68.3 &   0.0 \\
28.1 & 5805.23 & 4.17 &  -0.60 &  40.0 &  63.3 &   0.0 &  78.2 &   0.0 &   0.0 &   0.0 &  89.0 &  75.7 &   0.0 \\
28.1 & 6128.98 & 1.68 &  -3.39 & 118.9 &  99.0 & 131.0 &   0.0 &  96.5 & 143.5 &  97.1 & 124.6 &  92.4 & 128.2 \\
28.1 & 6130.14 & 4.26 &  -0.98 &  39.2 &  27.9 &  47.6 &   0.0 &  37.6 &  33.9 &  39.4 &  54.3 &  34.3 &  47.1 \\
28.1 & 6176.82 & 4.09 &  -0.24 &  81.8 & 106.0 & 118.3 &   0.0 &  84.2 & 110.3 &  75.7 & 103.6 &  61.9 &  97.7 \\
28.1 & 6177.25 & 1.83 &  -3.60 &  87.6 &  77.3 &  93.1 &   0.0 &  70.3 & 111.6 &  71.3 &  94.6 &  61.6 &  92.9 \\
28.1 & 6186.72 & 4.10 &  -0.90 &  68.2 &  47.5 &  57.3 &   0.0 &  60.7 &  68.4 &  54.0 &  73.3 &  50.3 &  74.0 \\
28.1 & 6204.61 & 4.09 &  -1.15 &   0.0 &  41.3 &  77.0 &   0.0 &  43.8 &  56.2 &  44.5 &  52.1 &  33.8 &  25.7 \\
28.1 & 6223.99 & 4.10 &  -0.97 &  46.2 &  42.5 &  54.1 &   0.0 &  47.0 &  49.6 &  39.9 &  51.9 &  34.4 &   0.0 \\
28.1 & 6230.10 & 4.10 &  -1.20 &  50.1 &  61.2 & 116.2 &   0.0 &  44.8 &  66.1 &  50.2 &  62.2 &  38.0 &  54.4 \\
28.1 & 6322.17 & 4.15 &  -1.21 &  32.2 &   0.0 &  26.0 &   0.0 &  35.3 &  28.4 &  24.3 &   0.0 &  23.3 &  33.6 \\
28.1 & 6327.60 & 1.68 &  -3.08 & 136.0 & 115.5 & 148.1 &   0.0 &  91.5 & 153.1 & 147.9 & 142.1 & 137.7 & 126.5 \\
28.1 & 6378.26 & 4.15 &  -0.82 &  57.9 &  49.8 &  66.9 &   0.0 &  52.4 &  55.0 &  64.7 &  63.6 &  55.6 &  51.4 \\
28.1 & 6384.67 & 4.15 &  -1.00 &  48.5 &  44.3 &  66.2 &   0.0 &  38.5 &  55.8 &  65.6 &  63.1 &  47.5 &  50.2 \\
56.1 & 6141.75 & 0.70 &   0.00 & 206.2 & 189.2 & 382.1 &   0.0 & 160.5 & 238.2 & 192.8 & 188.0 & 185.2 & 213.3 \\
\hline
\end{tabular}
\normalsize
\end{table*}

\addtocounter{table}{-1}

\begin{table*}
\caption{Equivalent Widths from Giraffe spectra (in electronic form)}
\tiny
\begin{tabular}{lrrrrrrrrrrrrr}
\hline
El   & Wavel.  & E.P. & log gf &7006377&7006470&7006590&7006935&7006983&7007064&7007118&7007884&7008891&7013582\\
     & (\AA)   & (eV) &        &(m\AA)~~&(m\AA)~~&(m\AA)~~&(m\AA)~~&(m\AA)~~&(m\AA)~~&(m\AA)~~&(m\AA)~~&(m\AA)~~&(m\AA)~~\\
\hline
 8.1 & 6300.31 & 0.00 &  -9.75 &  30.0 &   0.0 &  42.0 &  68.0 &  52.0 &  67.0 &  63.0 &  68.0 &  75.0 &  53.0 \\
 8.1 & 6363.79 & 0.02 & -10.25 &  23.0 &  26.0 &  23.0 &  35.0 &  37.0 &  30.3 &  30.7 &  34.0 &  34.0 &  32.0 \\
11.1 & 6154.23 & 2.10 &  -1.57 & 123.7 &  78.8 &  56.3 &  79.8 & 101.6 &  89.7 &  98.3 &  90.9 &  77.9 & 107.0 \\
11.1 & 6160.75 & 2.10 &  -1.26 & 148.2 & 104.6 &  79.0 & 118.6 & 116.7 & 101.3 & 126.1 & 114.6 & 100.5 & 124.6 \\
12.1 & 5711.09 & 4.34 &  -1.71 & 134.0 & 125.9 & 111.7 & 150.5 & 156.0 & 151.0 & 146.8 &   0.0 &   0.0 &   0.0 \\
12.1 & 6318.71 & 5.11 &  -1.97 &  75.0 &  84.9 &  77.8 &  85.2 &  88.5 &  77.9 &  93.8 &  94.2 &  79.3 &  93.1 \\
12.1 & 6319.24 & 5.11 &  -2.20 &  65.8 &  69.2 &  75.6 &  71.9 &  77.7 &  66.0 &  74.2 &  71.0 &  57.4 &  80.2 \\
14.1 & 5645.62 & 4.93 &  -2.14 &  63.3 &  60.6 &  60.1 &  55.6 &  64.5 &  58.7 &  65.5 &   0.0 &   0.0 &   0.0 \\
14.1 & 5665.56 & 4.92 &  -2.04 &  69.1 &  51.3 &  60.8 &  57.8 &  58.0 &  55.6 &  70.7 &   0.0 &   0.0 &   0.0 \\
14.1 & 5684.49 & 4.95 &  -1.65 &  66.9 &  53.3 &  66.8 &  56.8 &  65.9 &  56.2 &  69.5 &   0.0 &   0.0 &   0.0 \\
14.1 & 5690.43 & 4.93 &  -1.87 &  56.6 &  47.3 &  45.8 &  54.0 &  62.3 &  55.5 &  53.7 &   0.0 &   0.0 &   0.0 \\
14.1 & 5793.08 & 4.93 &  -2.06 &  28.5 &  56.3 &  50.2 &  61.7 &  56.1 &  61.1 &  57.6 &   0.0 &   0.0 &   0.0 \\
14.1 & 6125.03 & 5.61 &  -1.57 &  54.2 &  38.6 &  24.5 &  34.0 &  53.5 &  46.1 &  48.6 &  37.2 &  25.3 &  53.2 \\
14.1 & 6145.02 & 5.61 &  -1.44 &  48.8 &  35.8 &   0.0 &  30.0 &  54.6 &  40.1 &  45.3 &   0.0 &  22.3 &  52.0 \\
20.1 & 6161.30 & 2.52 &  -1.27 & 156.8 & 163.2 & 117.6 & 157.7 & 153.3 & 145.2 & 165.5 & 149.9 & 148.9 & 164.4 \\
20.1 & 6166.44 & 2.52 &  -1.14 & 146.5 & 131.9 &  96.4 & 142.3 & 143.4 & 138.1 & 152.4 & 131.6 & 140.5 & 148.6 \\
20.1 & 6169.04 & 2.52 &  -0.80 & 173.4 & 165.8 & 127.9 & 173.4 & 172.7 & 165.7 & 180.2 & 164.2 & 169.8 & 181.7 \\
20.1 & 6169.56 & 2.52 &  -0.48 & 188.8 & 167.7 & 155.0 & 192.1 & 189.4 & 164.5 & 186.3 & 171.9 & 189.1 & 189.5 \\
21.2 & 5640.99 & 1.50 &  -0.86 & 109.4 &   0.0 &  89.0 &  95.9 &  98.3 &  84.3 & 111.6 &   0.0 &   0.0 &   0.0 \\
21.2 & 5667.15 & 1.50 &  -1.11 &  73.5 &  46.1 &  96.2 &  87.0 &  98.7 &  94.6 & 110.8 &   0.0 &   0.0 &   0.0 \\
21.2 & 5669.04 & 1.50 &  -1.00 &  87.9 &  66.3 &  69.4 &  82.5 &  78.2 &  84.1 &  81.4 &   0.0 &   0.0 &   0.0 \\
21.2 & 5671.83 & 1.45 &   0.56 & 111.1 &  82.4 &  77.8 &  98.5 &  92.8 & 108.9 & 107.1 &   0.0 &   0.0 &   0.0 \\
21.2 & 6245.62 & 1.51 &  -1.05 & 100.0 & 106.3 &  85.0 & 110.7 &  96.0 &  85.8 & 100.0 & 108.9 & 109.3 &  83.5 \\
21.2 & 6279.74 & 1.50 &  -1.16 &  86.2 &  93.9 &  87.2 & 109.7 &  88.4 &  66.3 & 101.7 &  78.3 & 101.8 &  96.8 \\
22.1 & 6126.22 & 1.07 &  -1.42 & 161.5 & 149.2 & 114.4 & 174.3 & 158.7 & 176.1 & 176.4 & 160.3 & 158.6 & 125.3 \\
23.1 & 5624.89 & 1.05 &  -1.07 &  92.4 &  62.2 &  91.1 &  85.1 &  83.0 &  91.4 &  90.9 &   0.0 &   0.0 &   0.0 \\
23.1 & 5626.01 & 1.04 &  -1.25 & 111.6 &  72.9 &  93.6 & 113.5 &  93.6 & 111.5 & 125.5 &   0.0 &   0.0 &   0.0 \\
23.1 & 5627.64 & 1.08 &  -0.37 & 133.9 & 108.2 & 110.4 & 127.5 & 119.3 & 136.3 & 139.1 &   0.0 &   0.0 &   0.0 \\
23.1 & 5657.44 & 1.06 &  -1.02 & 100.9 &   0.0 &  68.7 &  93.4 &  85.8 & 105.9 &  96.8 &   0.0 &   0.0 &   0.0 \\
23.1 & 5668.36 & 1.08 &  -1.02 &   0.0 &  80.0 &  74.9 & 114.4 &  90.0 & 108.6 & 117.4 &   0.0 &   0.0 &   0.0 \\
23.1 & 5670.86 & 1.08 &  -0.42 & 130.7 & 107.9 & 103.8 & 135.9 & 118.5 & 145.0 & 138.7 &   0.0 &   0.0 &   0.0 \\
23.1 & 5698.53 & 1.06 &  -0.11 & 154.1 &   0.0 & 118.4 & 162.3 & 137.1 & 160.2 & 118.7 &   0.0 &   0.0 &   0.0 \\
23.1 & 5727.06 & 1.08 &  -0.01 & 166.6 &   0.0 & 129.9 & 173.5 & 160.1 & 180.7 & 167.8 &   0.0 &   0.0 &   0.0 \\
23.1 & 5737.06 & 1.06 &  -0.74 & 121.9 &   0.0 &  90.0 & 150.2 & 128.4 & 146.6 & 137.3 &   0.0 &   0.0 &   0.0 \\
23.1 & 5743.45 & 1.08 &  -0.97 & 108.7 & 104.3 &  92.2 & 134.9 & 103.5 & 134.1 & 121.6 &   0.0 &   0.0 &   0.0 \\
23.1 & 6251.82 & 0.29 &  -1.34 & 180.8 &   0.0 & 149.9 & 203.0 & 173.8 & 207.8 & 184.7 & 198.6 & 193.8 & 182.6 \\
24.1 & 5628.65 & 3.42 &  -0.77 &  49.3 &  43.9 &  62.3 &  51.3 &  33.1 &  50.6 &  66.5 &   0.0 &   0.0 &   0.0 \\
24.1 & 5783.07 & 3.32 &  -0.40 &  63.2 &  64.4 &  58.1 &  77.9 &  59.9 &  75.7 &  64.2 &   0.0 &   0.0 &   0.0 \\
24.1 & 5787.93 & 3.32 &  -0.08 &  86.2 &  91.2 &  76.5 & 102.9 &  95.2 & 119.4 &  98.8 &   0.0 &   0.0 &   0.0 \\
24.1 & 6330.10 & 0.94 &  -2.87 &  47.1 & 133.1 & 125.8 & 171.8 & 139.0 & 181.0 & 156.3 & 159.7 & 148.5 & 144.5 \\
26.1 & 5619.61 & 4.39 &  -1.49 &  60.5 &  54.7 &  55.8 &  63.0 &  63.1 &  61.0 &  71.9 &   0.0 &   0.0 &   0.0 \\
26.1 & 5635.83 & 4.26 &  -1.59 &  74.8 &  59.7 &  62.5 &  62.5 &  71.2 &  69.8 &  41.9 &   0.0 &   0.0 &   0.0 \\
26.1 & 5636.70 & 3.64 &  -2.53 &  76.6 &  60.5 &  63.0 &  65.5 &  61.9 &  52.5 &  79.9 &   0.0 &   0.0 &   0.0 \\
26.1 & 5650.00 & 5.10 &  -0.80 &  64.5 &  51.7 &  65.5 &  63.5 &  54.9 &  60.9 &  62.3 &   0.0 &   0.0 &   0.0 \\
26.1 & 5651.48 & 4.47 &  -1.79 &  40.0 &  41.0 &  44.7 &  47.0 &  46.7 &  46.9 &  46.7 &   0.0 &   0.0 &   0.0 \\
26.1 & 5652.33 & 4.26 &  -1.77 &  53.9 &  52.9 &  59.7 &  37.0 &  47.6 &  52.6 &  43.2 &   0.0 &   0.0 &   0.0 \\
26.1 & 5717.84 & 4.28 &  -0.98 &  91.8 &  90.2 &  76.8 & 105.5 & 104.1 &  91.5 &  97.1 &   0.0 &   0.0 &   0.0 \\
26.1 & 5731.77 & 4.26 &  -1.10 &  99.4 &  85.3 &  89.2 &  94.2 &  95.2 &  93.2 & 104.0 &   0.0 &   0.0 &   0.0 \\
26.1 & 5738.24 & 4.22 &  -2.24 &  51.3 &  41.9 &  41.6 &  51.7 &  47.2 &  42.8 &  58.3 &   0.0 &   0.0 &   0.0 \\
26.1 & 5741.86 & 4.26 &  -1.69 &  70.5 &  82.6 &  67.1 &  76.6 &  80.4 &  64.2 &  68.7 &   0.0 &   0.0 &   0.0 \\
26.1 & 5752.04 & 4.55 &  -0.92 &  85.2 &  82.8 &  67.7 &  94.7 &  80.3 &  68.4 &  84.2 &   0.0 &   0.0 &   0.0 \\
26.1 & 5775.09 & 4.22 &  -1.11 &  81.2 &  81.0 &  67.0 &  93.6 &  78.4 &  94.5 &  90.7 &   0.0 &   0.0 &   0.0 \\
26.1 & 5793.92 & 4.22 &  -1.62 &  80.6 &  68.4 &  64.5 &  68.8 &  67.0 &  77.4 &  63.8 &   0.0 &   0.0 &   0.0 \\
26.1 & 5806.73 & 4.61 &  -0.93 &  79.0 &  78.0 &  64.0 &  93.4 &  85.0 & 102.8 &  93.9 &   0.0 &   0.0 &   0.0 \\
26.1 & 5827.87 & 3.28 &  -3.16 &  65.7 &  61.1 &  47.8 &  75.2 &  58.0 &  86.6 &  62.3 &   0.0 &   0.0 &   0.0 \\
26.1 & 6151.62 & 2.18 &  -3.26 & 160.8 & 139.3 & 110.8 & 149.8 & 140.6 & 138.8 & 157.2 & 135.6 & 141.9 & 150.6 \\
26.1 & 6159.38 & 4.61 &  -1.88 &  39.2 &  35.6 &  23.3 &  29.4 &  40.8 &  43.3 &  29.0 &  25.1 &   0.0 &  45.0 \\
26.1 & 6165.36 & 4.14 &  -1.48 &  91.9 &  78.2 &  55.6 &  64.7 &  92.2 &  61.0 &  84.3 &  67.4 &  73.8 &  89.9 \\
26.1 & 6173.34 & 2.22 &  -2.84 & 173.5 & 166.6 & 128.2 & 168.1 & 156.2 & 156.1 & 169.4 & 165.9 & 160.5 & 159.0 \\
26.1 & 6187.40 & 2.83 &  -4.12 &  49.1 &  50.0 &  27.3 &   0.0 &  38.4 &  46.8 &  43.3 &  33.4 &   0.0 &  30.3 \\
26.1 & 6188.00 & 3.94 &  -1.60 &  94.6 & 103.1 &  61.2 &  87.7 & 103.2 &  85.3 & 105.7 &  75.3 &  86.3 &  92.2 \\
26.1 & 6200.32 & 2.61 &  -2.39 & 161.9 & 141.3 & 117.0 & 158.5 & 150.4 & 140.5 & 161.3 & 158.2 & 167.5 & 148.9 \\
26.1 & 6213.44 & 2.22 &  -2.54 & 192.0 & 164.4 & 161.0 & 200.6 & 169.1 & 185.6 & 183.8 & 185.6 & 181.5 & 161.3 \\
26.1 & 6219.29 & 2.20 &  -2.39 & 197.8 & 176.2 & 166.6 & 209.5 &   0.0 & 199.9 & 186.7 & 200.4 & 190.8 & 164.4 \\
26.1 & 6232.65 & 3.65 &  -1.21 & 148.6 & 140.1 & 125.9 & 154.3 &   0.0 & 139.1 & 149.8 & 145.7 & 148.8 & 138.7 \\
26.1 & 6240.65 & 2.22 &  -3.23 & 142.0 & 142.7 & 136.2 & 159.0 & 128.9 & 136.4 & 144.3 & 142.0 & 144.8 & 137.2 \\
26.1 & 6265.14 & 2.18 &  -2.51 & 208.5 & 158.2 & 174.6 & 218.9 & 174.3 & 216.1 & 193.2 & 218.7 & 199.3 & 178.4 \\
26.1 & 6270.23 & 2.86 &  -2.55 & 118.2 & 122.1 & 104.5 & 138.6 & 125.1 &   0.0 & 121.6 & 133.5 &  43.3 & 111.7 \\
26.1 & 6311.50 & 2.83 &  -3.16 &  95.9 &  96.4 &  93.9 & 115.8 &  97.1 & 105.4 & 101.7 & 126.5 & 111.1 &  96.2 \\
26.1 & 6315.81 & 4.07 &  -1.67 &  59.8 &  83.2 &  86.5 &  81.5 &  77.3 &  43.1 &  86.9 &  81.6 &  73.1 &  76.0 \\
26.1 & 6322.69 & 2.59 &  -2.38 & 157.7 & 136.0 & 141.5 & 165.0 & 143.7 & 170.5 & 149.5 & 172.0 & 153.3 & 135.7 \\
26.1 & 6330.85 & 4.73 &  -1.22 &  55.2 &  58.6 &  53.1 &  60.8 &  71.0 &  56.2 &  70.5 &  56.5 &  56.8 &  69.0 \\
26.1 & 6335.34 & 2.20 &  -2.28 & 218.7 & 169.6 & 196.6 & 188.0 & 105.9 & 205.2 & 211.9 & 216.0 & 192.9 & 197.1 \\
26.1 & 6353.84 & 0.91 &  -6.41 &  63.0 &  44.0 &  63.3 &  70.5 &  48.6 &  60.1 &  64.3 &  74.2 &  56.0 &  59.2 \\
26.1 & 6380.75 & 4.19 &  -1.34 &  97.5 &  56.2 &  82.7 &  99.5 &  79.9 &  98.6 &  87.3 & 106.6 &  83.3 &  86.6 \\
26.1 & 6392.54 & 2.28 &  -3.97 &   0.0 &  72.0 &  75.3 &  85.7 &  69.8 & 110.2 &  79.4 &  99.1 &  74.5 &  75.5 \\
26.2 & 6247.56 & 3.87 &  -2.32 &  40.1 &  35.7 &   0.0 &  36.2 &   0.0 &  33.3 &  42.2 &  33.8 &  43.1 &  25.7 \\
26.2 & 6369.46 & 2.89 &  -4.21 &   0.0 &  36.4 &  41.0 &  25.0 &  31.5 &  34.3 &  34.6 &  34.3 &  29.0 &  33.7 \\
28.1 & 5643.09 & 4.16 &  -1.25 &  35.3 &  33.3 &  31.6 &  36.7 &  34.2 &  40.4 &  38.6 &   0.0 &   0.0 &   0.0 \\
28.1 & 5709.56 & 1.68 &  -2.14 & 151.3 & 147.6 & 143.2 & 166.7 & 148.2 & 163.8 & 152.9 &   0.0 &   0.0 &   0.0 \\
28.1 & 5760.84 & 4.10 &  -0.81 &  65.3 &  82.8 &  60.0 &  71.2 &  65.5 &  65.9 &  76.6 &   0.0 &   0.0 &   0.0 \\
28.1 & 5805.23 & 4.17 &  -0.60 &  67.0 &  70.9 &  63.7 &  73.7 &  76.5 & 100.1 &  75.8 &   0.0 &   0.0 &   0.0 \\
28.1 & 6128.98 & 1.68 &  -3.39 & 136.7 & 118.5 &  76.4 & 123.1 & 116.1 & 123.8 & 136.3 & 110.8 & 112.1 & 124.1 \\
28.1 & 6130.14 & 4.26 &  -0.98 &  45.1 &  46.5 &  28.1 &  33.9 &  46.8 &  43.0 &  48.3 &  29.4 &  22.3 &  46.3 \\
28.1 & 6176.82 & 4.09 &  -0.24 & 108.9 & 102.9 &  66.6 &  86.4 &  98.3 &  93.7 & 111.4 &  67.3 &  93.4 &  99.6 \\
28.1 & 6177.25 & 1.83 &  -3.60 &  95.6 &  86.4 &  61.5 &  79.9 &  89.7 &  91.5 &  96.6 &  66.3 &  71.6 &  89.7 \\
28.1 & 6186.72 & 4.10 &  -0.90 &  63.0 &  77.3 &  42.0 &  47.3 &  74.8 &  74.1 &  74.1 &  48.7 &  51.2 &  77.5 \\
28.1 & 6204.61 & 4.09 &  -1.15 &  52.3 &  53.0 &  28.0 &  43.5 &  55.3 &  44.4 &  55.9 &  43.9 &  49.6 &  64.6 \\
28.1 & 6223.99 & 4.10 &  -0.97 &  49.3 &  34.1 &  31.5 &  42.5 &   0.0 &  49.9 &  41.5 &  30.2 &  46.3 &   0.0 \\
28.1 & 6230.10 & 4.10 &  -1.20 &  56.3 &  68.7 &  33.5 &  54.9 &   0.0 &  50.8 &  55.5 &  47.3 &  54.2 &  63.2 \\
28.1 & 6322.17 & 4.15 &  -1.21 &  38.0 &  39.6 &  26.3 &  26.9 &  44.5 &  25.0 &  34.3 &  33.5 &  34.3 &  37.2 \\
28.1 & 6327.60 & 1.68 &  -3.08 & 146.9 & 140.8 & 129.1 & 157.6 & 132.0 & 152.8 & 141.6 & 169.7 & 142.9 & 137.5 \\
28.1 & 6378.26 & 4.15 &  -0.82 &  54.8 &   0.0 &  55.0 &  60.4 &  54.2 &  71.2 &  55.2 &  75.2 &  52.4 &  64.3 \\
28.1 & 6384.67 & 4.15 &  -1.00 &  51.2 &  46.6 &  48.2 &  59.5 &   0.0 &  48.4 &  53.4 &  66.8 &  47.8 &  50.8 \\
56.1 & 6141.75 & 0.70 &   0.00 & 209.5 & 186.5 & 156.7 & 209.9 & 189.6 & 211.5 & 207.8 & 222.1 & 210.5 & 184.3 \\
\hline
\end{tabular}
\label{t:ewidths}
\normalsize
\end{table*}

\addtocounter{table}{-1}

\begin{table*}
\caption{Equivalent Widths from Giraffe spectra (in electronic form)}
\tiny
\begin{tabular}{lrrrrrrrr}
\hline
El   & Wavel.  & E.P. & log gf &8005158&8005404&8006535&8008693&8013657\\
     & (\AA)   & (eV) &        &(m\AA)~~&(m\AA)~~&(m\AA)~~&(m\AA)~~&(m\AA)~~\\
\hline
 8.1 & 6300.31 & 0.00 &  -9.75 &  58.0 &  82.0 &  54.0 &  50.0 &  55.4 \\
 8.1 & 6363.79 & 0.02 & -10.25 &  34.0 &  33.0 &  33.0 &  19.0 &  23.7 \\
11.1 & 6154.23 & 2.10 &  -1.57 &  92.7 &  80.3 &  94.1 & 165.0 &  86.1 \\
11.1 & 6160.75 & 2.10 &  -1.26 & 122.6 & 117.5 & 129.4 & 174.1 & 116.6 \\
12.1 & 5711.09 & 4.34 &  -1.71 &   0.0 & 171.7 & 139.0 &   0.0 &   0.0 \\
12.1 & 6318.71 & 5.11 &  -1.97 &  90.0 &  87.4 & 102.1 &   0.0 &  64.1 \\
12.1 & 6319.24 & 5.11 &  -2.20 &  67.5 &  75.1 &  90.7 &   0.0 &  56.3 \\
14.1 & 5645.62 & 4.93 &  -2.14 &   0.0 &  74.9 &  75.5 &  72.8 &   0.0 \\
14.1 & 5665.56 & 4.92 &  -2.04 &   0.0 &  59.9 &  67.8 &   0.0 &   0.0 \\
14.1 & 5684.49 & 4.95 &  -1.65 &   0.0 & 102.7 &  78.8 &  41.1 &   0.0 \\
14.1 & 5690.43 & 4.93 &  -1.87 &   0.0 &  76.7 &  60.7 &  45.1 &   0.0 \\
14.1 & 5793.08 & 4.93 &  -2.06 &   0.0 &  54.1 &  59.5 &  70.5 &   0.0 \\
14.1 & 6125.03 & 5.61 &  -1.57 &  52.9 &  31.0 &  48.7 &  94.3 &  55.8 \\
14.1 & 6145.02 & 5.61 &  -1.44 &  52.1 &  28.8 &  38.4 &   0.0 &  57.9 \\
20.1 & 6161.30 & 2.52 &  -1.27 & 164.4 & 163.9 & 172.8 & 192.5 & 121.7 \\
20.1 & 6166.44 & 2.52 &  -1.14 & 151.3 & 155.7 & 140.6 & 164.7 & 124.9 \\
20.1 & 6169.04 & 2.52 &  -0.80 & 168.8 & 177.6 & 187.2 & 201.6 & 144.5 \\
20.1 & 6169.56 & 2.52 &  -0.48 & 155.8 & 194.5 & 197.0 & 207.8 & 145.8 \\
21.2 & 5640.99 & 1.50 &  -0.86 &   0.0 & 125.3 & 110.0 &  94.1 &   0.0 \\
21.2 & 5667.15 & 1.50 &  -1.11 &   0.0 & 102.0 & 106.3 &  79.8 &   0.0 \\
21.2 & 5669.04 & 1.50 &  -1.00 &   0.0 &  21.1 & 103.3 &  93.2 &   0.0 \\
21.2 & 5671.83 & 1.45 &   0.56 &   0.0 & 152.6 & 145.7 &  81.1 &   0.0 \\
21.2 & 6245.62 & 1.51 &  -1.05 &  96.6 & 105.8 & 110.1 &  99.5 &  79.9 \\
21.2 & 6279.74 & 1.50 &  -1.16 &  73.1 &  64.4 &  96.5 &  91.9 &  50.5 \\
22.1 & 6126.22 & 1.07 &  -1.42 & 160.1 & 172.1 & 179.9 & 191.3 & 128.7 \\
23.1 & 5624.89 & 1.05 &  -1.07 &   0.0 & 137.7 & 114.1 &  92.3 &   0.0 \\
23.1 & 5626.01 & 1.04 &  -1.25 &   0.0 & 132.0 & 140.9 &  80.1 &   0.0 \\
23.1 & 5627.64 & 1.08 &  -0.37 &   0.0 & 173.0 & 146.9 &  83.7 &   0.0 \\
23.1 & 5657.44 & 1.06 &  -1.02 &   0.0 & 114.9 & 102.8 &  65.8 &   0.0 \\
23.1 & 5668.36 & 1.08 &  -1.02 &   0.0 & 111.0 & 130.3 &  69.2 &   0.0 \\
23.1 & 5670.86 & 1.08 &  -0.42 &   0.0 & 193.5 & 163.0 &  91.0 &   0.0 \\
23.1 & 5698.53 & 1.06 &  -0.11 &   0.0 & 234.6 & 183.3 & 150.4 &   0.0 \\
23.1 & 5727.06 & 1.08 &  -0.01 &   0.0 & 208.5 & 167.3 &  87.0 &   0.0 \\
23.1 & 5737.06 & 1.06 &  -0.74 &   0.0 & 140.3 & 138.2 & 107.3 &   0.0 \\
23.1 & 5743.45 & 1.08 &  -0.97 &   0.0 & 125.9 & 132.4 &  69.4 &   0.0 \\
23.1 & 6251.82 & 0.29 &  -1.34 & 162.1 & 190.7 & 217.7 &  58.5 & 146.8 \\
24.1 & 5628.65 & 3.42 &  -0.77 &   0.0 &  68.8 &  55.9 &  43.7 &   0.0 \\
24.1 & 5783.07 & 3.32 &  -0.40 &   0.0 &  86.1 &  66.1 & 104.1 &   0.0 \\
24.1 & 5787.93 & 3.32 &  -0.08 &   0.0 &  94.5 &  92.7 & 138.8 &   0.0 \\
24.1 & 6330.10 & 0.94 &  -2.87 & 141.3 & 167.9 & 192.7 & 198.4 & 114.5 \\
26.1 & 5619.61 & 4.39 &  -1.49 &   0.0 &  79.6 &  64.5 &  51.2 &   0.0 \\
26.1 & 5635.83 & 4.26 &  -1.59 &   0.0 &  78.9 &  87.1 &  63.8 &   0.0 \\
26.1 & 5636.70 & 3.64 &  -2.53 &   0.0 &  70.4 &  66.9 &  72.1 &   0.0 \\
26.1 & 5650.00 & 5.10 &  -0.80 &   0.0 &  59.5 &  77.0 &  39.0 &   0.0 \\
26.1 & 5651.48 & 4.47 &  -1.79 &   0.0 &  43.8 &  56.7 &  50.3 &   0.0 \\
26.1 & 5652.33 & 4.26 &  -1.77 &   0.0 &  54.6 &  61.6 &  61.6 &   0.0 \\
26.1 & 5717.84 & 4.28 &  -0.98 &   0.0 & 111.6 &  92.5 &  77.3 &   0.0 \\
26.1 & 5731.77 & 4.26 &  -1.10 &   0.0 &  98.2 &  97.4 &  64.9 &   0.0 \\
26.1 & 5738.24 & 4.22 &  -2.24 &   0.0 &  29.3 &  51.2 &  64.8 &   0.0 \\
26.1 & 5741.86 & 4.26 &  -1.69 &   0.0 &  63.6 &  71.5 &   0.0 &   0.0 \\
26.1 & 5752.04 & 4.55 &  -0.92 &   0.0 &  64.0 &  70.3 &  54.1 &   0.0 \\
26.1 & 5775.09 & 4.22 &  -1.11 &   0.0 &  90.5 &  89.5 &  56.3 &   0.0 \\
26.1 & 5793.92 & 4.22 &  -1.62 &   0.0 &  57.1 &  72.5 &  52.7 &   0.0 \\
26.1 & 5806.73 & 4.61 &  -0.93 &   0.0 &  62.6 &  62.5 &  55.5 &   0.0 \\
26.1 & 5827.87 & 3.28 &  -3.16 &   0.0 &  33.2 &  70.2 &  60.0 &   0.0 \\
26.1 & 6151.62 & 2.18 &  -3.26 & 146.4 & 136.6 & 143.3 & 168.7 & 102.5 \\
26.1 & 6159.38 & 4.61 &  -1.88 &  40.7 &  22.8 &  49.9 &  88.9 &  36.8 \\
26.1 & 6165.36 & 4.14 &  -1.48 &  84.1 &  64.0 &  64.1 & 100.1 &  81.3 \\
26.1 & 6173.34 & 2.22 &  -2.84 & 161.3 & 154.8 & 164.4 & 172.1 & 126.9 \\
26.1 & 6187.40 & 2.83 &  -4.12 &  41.0 &  30.1 &  45.5 &  36.8 &  36.6 \\
26.1 & 6188.00 & 3.94 &  -1.60 &  95.2 &  76.1 &  81.3 & 122.0 &  77.0 \\
26.1 & 6200.32 & 2.61 &  -2.39 & 149.1 & 152.9 & 135.0 & 147.4 & 118.6 \\
26.1 & 6213.44 & 2.22 &  -2.54 & 176.5 & 187.1 & 195.7 & 175.3 & 140.4 \\
26.1 & 6219.29 & 2.20 &  -2.39 & 182.9 & 186.9 & 200.4 & 197.5 & 149.8 \\
26.1 & 6232.65 & 3.65 &  -1.21 & 143.8 & 110.2 & 148.5 & 159.8 & 109.1 \\
26.1 & 6240.65 & 2.22 &  -3.23 & 125.1 & 147.6 & 155.6 & 139.6 & 107.5 \\
26.1 & 6265.14 & 2.18 &  -2.51 & 175.8 & 183.2 & 217.3 & 239.2 & 164.9 \\
26.1 & 6270.23 & 2.86 &  -2.55 & 124.8 & 121.9 & 133.0 &  27.3 &  96.9 \\
26.1 & 6311.50 & 2.83 &  -3.16 &  92.7 & 103.9 & 121.8 & 128.6 &  66.2 \\
26.1 & 6315.81 & 4.07 &  -1.67 &   0.0 &  73.0 &  82.0 &  90.7 &  48.5 \\
26.1 & 6322.69 & 2.59 &  -2.38 & 143.3 & 148.3 & 169.9 & 171.9 & 113.8 \\
26.1 & 6330.85 & 4.73 &  -1.22 &  40.9 &  66.4 &  60.2 & 100.9 &  48.3 \\
26.1 & 6335.34 & 2.20 &  -2.28 & 178.1 & 190.8 & 238.5 & 253.9 & 165.3 \\
26.1 & 6353.84 & 0.91 &  -6.41 &  44.5 &  53.7 &  83.7 & 100.5 &  46.1 \\
26.1 & 6380.75 & 4.19 &  -1.34 &  85.6 &  88.0 & 116.4 & 123.3 &  87.7 \\
26.1 & 6392.54 & 2.28 &  -3.97 &  67.2 &  75.6 & 101.0 & 106.2 &  78.8 \\
26.2 & 6247.56 & 3.87 &  -2.32 &  69.0 &  40.4 &  25.1 &  53.3 &  29.0 \\
26.2 & 6369.46 & 2.89 &  -4.21 &  34.5 &  31.2 &  31.9 &  51.3 &  32.0 \\
28.1 & 5643.09 & 4.16 &  -1.25 &   0.0 &  22.8 &  36.6 &  73.9 &   0.0 \\
28.1 & 5709.56 & 1.68 &  -2.14 &   0.0 & 257.1 & 172.7 &  21.1 &   0.0 \\
28.1 & 5760.84 & 4.10 &  -0.81 &   0.0 &  52.8 &  63.6 &  59.3 &   0.0 \\
28.1 & 5805.23 & 4.17 &  -0.60 &   0.0 &  50.9 &  76.2 &  53.5 &   0.0 \\
28.1 & 6128.98 & 1.68 &  -3.39 & 121.3 & 106.1 & 116.9 & 163.1 & 109.6 \\
28.1 & 6130.14 & 4.26 &  -0.98 &  43.2 &  27.7 &  43.3 &  81.5 &  52.6 \\
28.1 & 6176.82 & 4.09 &  -0.24 &  98.9 &  81.4 &  84.6 & 121.3 &  81.7 \\
28.1 & 6177.25 & 1.83 &  -3.60 &  86.7 &  70.0 &  80.1 & 125.1 &  82.7 \\
28.1 & 6186.72 & 4.10 &  -0.90 &  64.2 &  45.9 &  68.8 & 127.1 &  55.8 \\
28.1 & 6204.61 & 4.09 &  -1.15 &  46.5 &  29.9 &  44.4 &  92.4 &  50.9 \\
28.1 & 6223.99 & 4.10 &  -0.97 &  45.7 &  40.0 &  46.6 &  82.9 &  41.0 \\
28.1 & 6230.10 & 4.10 &  -1.20 &  48.2 &  42.3 &  50.7 &  98.8 &  59.5 \\
28.1 & 6322.17 & 4.15 &  -1.21 &  33.1 &  24.0 &  31.3 &  54.4 &  24.4 \\
28.1 & 6327.60 & 1.68 &  -3.08 & 129.5 & 149.7 & 170.9 & 171.1 & 129.1 \\
28.1 & 6378.26 & 4.15 &  -0.82 &  55.5 &  62.8 &  85.8 &  97.7 &  52.4 \\
28.1 & 6384.67 & 4.15 &  -1.00 &  49.4 &  45.9 &  72.2 &  97.7 &  42.3 \\
56.1 & 6141.75 & 0.70 &   0.00 & 194.8 & 205.2 & 210.2 & 213.9 & 178.2 \\
\hline
\end{tabular}
\normalsize
\end{table*}

\section{ABUNDANCE ANALYSIS}

\subsection{Equivalent Widths}


The equivalent widths (EWs) were measured using the
ROSA code (Gratton, private communication; see Table~\ref{t:ewidths} for stars members
of \object{NGC 6441}) with Gaussian fittings to the measured profiles: these
exploit a linear relation between EWs and FWHM of the lines, derived
from a subset of lines characterized by cleaner profiles (see
Bragaglia et al. 2001 for further details on this procedure). Since
the observed stars span a very limited atmospheric parameter range (
3908 $\leq T_{\rm eff} \leq$ 4321\,K and 
1.26 $\leq \log g \leq$ 1.74), internal
errors in these EWs may be estimated by comparing measures for
individual stars with the average values over the whole sample: the
values given in Column 10 of Table~\ref{t:phot} are the r.m.s. of
residuals around a best fit line, after eliminating a few
outliers. These errors may be slightly overestimated, due to real
star-to-star differences. They are roughly reproduced by the formula
$\sigma$(EW)$\sim 416/(S/N)$~m\AA. Considering the resolution and
sampling of the spectra, the errors in the EWs are about 1.3 times
larger than expectations based on photon noise statistics (Cayrel 1988
- considering that we used only the inner part of the profile
when deriving the equivalent widths), showing that a significant
contribution to errors is due to uncertainties in the correct
positioning of the continuum level; the observed errors could be
justified by errors of slightly less than 1\% in the estimate of the
correct continuum level.

\begin{figure}[h]
\includegraphics[width=8.8cm]{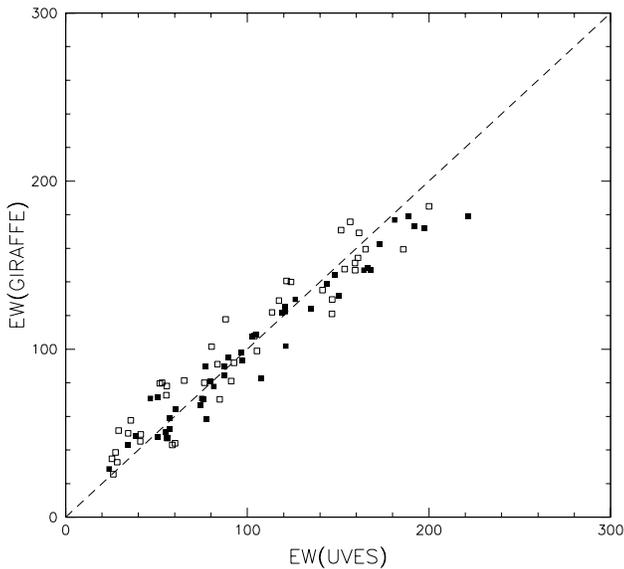}
\caption[]{Comparison between the EWs derived from Giraffe and UVES spectra for the two stars which
were observed in both modes. Filled symbols are for star \#6003734; open
symbols are for star \#7004329.}
\label{f:ewidths}
\end{figure}

Tests on possible systematic errors in the EWs may be done by
comparing them with those from other data sets. We could perform this
analysis for two stars (\#6003734 and \#7004329) having both UVES and
Giraffe spectra (taken on different observing set ups). Both stars are
not members of \object{NGC 6441}. On average the EWs measured on the Giraffe
spectra are larger than those measured on the UVES spectra by $0.3\pm
1.6$~m\AA, with an r.m.s. scatter of 14.5~m\AA\ (see Figure \ref{f:ewidths} for a
graphical comparison). 
The regression line through the points in  this figure 
is EW(GIRAFFE)=(0.84$\pm$0.03) EW(UVES)+ (16$\pm$ 12)\, m\AA.
While the scatter agrees with expected errors on both
sets of EWs, the regression line suggests that the EWs 
in the GIRAFFE spectra can be 
overestimated for the weaker lines and underestimated for the strong ones.
However, we deem this result too uncertain to apply any systematic correction to 
the present EWs. The impact of this potential error on the derived abundances 
will be commented later on in the text. 

\begin{table*}
\begin{center}
\caption{Atmospheric Parameters, Fe {\sc I},{\sc II}, Na and O abundances for stars observed with Giraffe}
\begin{tabular}{lcccccccccccc}
\hline
Star &T$_{\rm eff}$&$\log g$&$[$A/H$]$&$v_t$&\multicolumn{3}{c}{Fe I}&\multicolumn{3}{c}{Fe II}&[O/Fe]&[Na/Fe]\\
     &     (K)     &        &         &(km s$^{-1}$)&n&[Fe/H] &rms&n&[Fe/H] &rms&&\\
\hline
6005198 & 3908 & 1.26 &$-$0.32 & 1.4 & 32 &$-$0.41 & 0.23 &2&$-$0.29&0.52&$-$0.08 &   0.65 \\
6005934 & 4161 & 1.56 &$-$0.32 & 1.4 & 33 &$-$0.42 & 0.24 &2&$-$0.36&0.14&   0.08 &   0.17 \\
6007741 & 4199 & 1.60 &$-$0.32 & 1.7 & 20 &$-$0.25 & 0.26 &1&$-$0.95&    &$-$0.17 &   0.55 \\ 
6010149 & 4262 & 1.67 &$-$0.32 & 1.7 & 13 &$-$0.17 & 0.18 & &       &    &      &        \\ 
6012768 & 4304 & 1.73 &$-$0.32 & 1.0 & 19 &$-$0.42 & 0.20 &2&   0.00&0.75&$-$0.02 &   0.40 \\
7004955 & 3968 & 1.33 &$-$0.32 & 1.7 & 18 &$-$0.22 & 0.18 &1&$-$0.68&    &$-$0.39 &   1.01 \\
7006255 & 4162 & 1.56 &$-$0.32 & 1.6 & 34 &$-$0.39 & 0.19 &1&$-$0.98&    &$-$0.31 &   0.65 \\
7006305 & 4105 & 1.49 &$-$0.32 & 1.4 & 33 &$-$0.29 & 0.21 &2&$-$0.28&0.42&$-$0.05 &   0.70 \\
7006319 & 4136 & 1.53 &$-$0.32 & 1.4 & 34 &$-$0.29 & 0.22 &1&$-$0.06&    &   0.01 &   0.12 \\
7006354 & 4019 & 1.39 &$-$0.32 & 1.6 & 18 &$-$0.41 & 0.15 &2&$-$0.19&0.22&   0.04 &   0.40 \\
7006377 & 4085 & 1.47 &$-$0.32 & 1.8 & 33 &$-$0.37 & 0.16 &1&$-$0.44&    &$-$0.21 &   0.57 \\
7006470 & 4163 & 1.56 &$-$0.32 & 1.4 & 34 &$-$0.29 & 0.21 &2&$-$0.24&0.46&   0.11 &   0.18 \\
7006590 & 4247 & 1.66 &$-$0.32 & 1.4 & 34 &$-$0.41 & 0.20 &1&   0.13&    &   0.02 &$-$0.10 \\
7006935 & 4095 & 1.48 &$-$0.32 & 2.0 & 33 &$-$0.43 & 0.16 &2&$-$0.43&0.24&   0.15 &   0.07 \\
7006983 & 4178 & 1.58 &$-$0.32 & 1.4 & 32 &$-$0.13 & 0.17 &1&$-$0.06&    &   0.17 &   0.43 \\
7007064 & 4013 & 1.38 &$-$0.32 & 1.8 & 31 &$-$0.45 & 0.21 &2&$-$0.21&0.46&   0.05 &   0.03 \\
7007118 & 4209 & 1.61 &$-$0.32 & 1.8 & 34 &$-$0.35 & 0.18 &2&$-$0.31&0.33&   0.19 &   0.35 \\
7007884 & 4044 & 1.42 &$-$0.32 & 1.7 & 19 &$-$0.31 & 0.20 &2&$-$0.22&0.45&   0.12 &   0.16 \\
7008891 & 4290 & 1.71 &$-$0.32 & 1.7 & 18 &$-$0.30 & 0.17 &2&$-$0.42&0.22&   0.35 &   0.14 \\
7013582 & 4321 & 1.74 &$-$0.32 & 1.7 & 21 &$-$0.35 & 0.18 &2&$-$0.62&0.64&   0.22 &   0.50 \\
8005158 & 4081 & 1.46 &$-$0.32 & 1.6 & 19 &$-$0.40 & 0.18 &2&   0.19&0.17&   0.09 &   0.28 \\
8005404 & 3956 & 1.32 &$-$0.32 & 1.6 & 33 &$-$0.42 & 0.21 &2&$-$0.03&0.23&   0.13 &   0.10 \\
8006535 & 3983 & 1.35 &$-$0.32 & 1.6 & 33 &$-$0.25 & 0.21 &2&$-$0.30&0.59&$-$0.02 &   0.27 \\
8008693 & 4105 & 1.49 &$-$0.32 & 1.8 & 32 &$-$0.30 & 0.42 &2&   0.12&0.39&$-$0.10 &   1.01 \\
8013657 & 4308 & 1.73 &$-$0.32 & 1.0 & 21 &$-$0.27 & 0.17 &2&$-$0.45&0.54&   0.18 &   0.54 \\
\hline
\end{tabular}
\label{t:giraffeatmo}
\end{center}
\end{table*}

\subsection{Atmospheric Parameters}


We performed a standard line analysis on the equivalent widths
measured on our spectra, using model atmospheres extracted by
interpolation from the grid by Kurucz (1992; models with the
overshooting option switched off). Atmospheric parameters defining
these model atmospheres were obtained as follows.

Whenever possible, effective temperatures were derived  from a
calibration of $K$\ magnitudes, drawn from the 2MASS catalog (Skrutskie
et al. 2006). The idea is quite simple: our purpose is to derive the most
accurate temperatures from individual stars in a case where the dominant
source of error for temperatures derived from photometry is differential
reddening. We may accept a scale error (that is a zero point error common
to all stars), but we wish to reduce the star-to-star error. $K$\
magnitudes are much less affected by reddening than $V-K$\ colors. Hence,
if the dominant source of error is differential reddening, the best
solution would be to have a
relation where we enter $K$\ magnitudes and retrieve the corresponding
$T_{\rm eff}$\ values. This is
possible in a cluster insofar we may assume that there is a unique
relation between luminosity and effective temperature, which is a
reasonable assumption along the RGB if all cluster stars have the same
metallicity (see below).

 In principle, we would need a different $K-T_{\rm
eff}$\ relation for each star, because of differential reddening; however,
this individual $K-T_{\rm eff}$\ relation is very close to the average one
over all cluster stars, because the impact of differential reddening on
$K$\ magnitudes is very limited. In fact, roughly speaking,
$A(K)=0.4~E(B-V)$, hence for a differential reddening r.m.s. of 0.05~mag
(Layden et al. 1999; Pritzl et al. 2001), the different $K-T_{\rm eff}$\
relations differ from the average one by some 0.02~mag r.m.s.; since the
slope of the $K-T_{\rm eff}$\ relation is $332$~K/mag, the errors made by
adopting a unique average relation rather than one appropriate for each
star is $0.02\times 332=7$~K. The internal errors in the temperatures of
individual stars are actually larger, because the photometric errors in
the 2MASS $K$\ magnitudes ($\sim 0.03$~mag) should also be taken into
account: they cause an error of about 10~K in the adopted temperatures.
While the two errors should be summed in quadrature, we conservatively
simply added them and attribute an internal error of $\pm 17$~K to the
individual $T_{\rm eff}$'s.

The problem of deriving accurate temperatures then reduces to the
derivation of the average $K-T_{\rm eff}$\ relation for the cluster. This
is obtained by fitting a straight line through the $K-T_{\rm eff}(V-K)$\
points\footnote{The rms scatter around this straight line is of about 80 K, which is consistent
with a dispersion of 0.06 mag in the reddening values, to be compared with
the 0.05 mag proposed by Layden et al. (1999) and Pritzl et al. (2001), on a much larger
number of stars., and it is obviously valid only for that cluster. Each individual
$T_{\rm eff}(V-K)$\ is the temperature for each star derived from the
$V-K$\ color, de-reddened adopting the average cluster reddening (the
value we adopted is $E(B-V)=0.49$, which is the average value between
various literature determinations; see Paper II for a discussion; this was
transformed into $E(V-K)$\ using the formula $E(V-K)=2.75~E(B-V)$,
Cardelli et al. 1989), and using the calibration by Alonso et al. (1999;
we adopted the same average metal abundance for all stars when using
Alonso et al. formulas); the uncertainty in this average value translates
into a scale error, that is much larger than the error for the individual
stars because in this case we have to use the $V-K$\ color: since the
mean interstellar reddening toward NGC6441 is uncertain by about 0.05 mag
in $E(B-V)$, that is 0.13~mag in $E(V-K)$, and the slope of the $T_{\rm
eff}(V-K)$\ relation is 488~K/mag, the zero point of our $T_{\rm eff}$\
scale has an uncertainty of at least $\pm 67$~K. Note that there should
also be an additional statistical contribution, related to the spread of
differential reddening and the number of stars used, but this is
 practically negligible. This error bar does not include  possible errors 
in the Alonso et al. (1999) calibration. We will not consider here this
 last error, since we intend to adopt the Alonso et al. calibration
throughout the whole present series of papers.

Practically, a complication in this approach is that there are many field
interlopers. Luckily, radial velocities and positions can be used to eliminate most of
them. The relations we used are based on the bona fide cluster members
alone.

As mentioned above, a basic assumption behind this approach is that the
intrinsic width of the RGB of the cluster is negligible: this implies that
all stars share the same metal abundance. A consistency check can be
obtained a posteriori, looking at the spread in metal abundances. We
indeed found that the scatter in metallicities among cluster members is
consistent with expectations based on internal errors alone. Hence we have
no reason to think that there is a real spread in metal abundances among
the stars of NGC6441.}

For a few stars lacking of the 2MASS photometric data we inferred the $K$\
magnitudes by deriving the ($V-K$)\ color from our ($V-I$)\ colors. The
mean relation is ($V-K$)$=0.479+1.952~$($V-I$), derived from more than 250
stars spanning over 4 magnitudes in ($V-K$)\ colors; the r.m.s. scatter
around this mean relation is of 0.098 mag in ($V-K$). We could 
derive consistent temperatures also for these stars; in this case the
error bars are larger (about 40 K), since errors in their $K$\
magnitudes are of about 0.1~mag.

We may compare these temperatures derived from colors with those that
we could deduce from excitation equilibrium for Fe I lines. We found
that temperatures derived from Fe I excitation equilibrium
are lower by $41\pm 29$~K, with an r.m.s. scatter for individual stars
of 145~K on average. This small difference can be attributed to several causes
(errors in the adopted temperature scale, inadequacies of the adopted
model atmospheres, etc.). On the whole we do not deem this
difference as important. On the other hand, the fair agreement
between temperatures from colors and spectra supports the assumed
reddening (actually the best guess would be for a value of
$E(B-V)=0.459\pm 0.022$, slightly lower than the value adopted here).
  
Surface gravities were obtained from the location of the stars in the
CMD. This procedure requires assumptions about
the distance modulus (we adopted $(m-M)_V=16.33$\ from Harris 1996 for
the cluster members), the bolometric corrections (from Alonso et
al. 1999), and the masses (we assumed a mass of 0.9~M$_\odot$, close
to the value given by isochrone fitting). Uncertainties in these
gravities are small for cluster stars (we estimate internal
star-to-star errors of about 0.02~dex, due to the effects of possible
variations in the interstellar absorption of 0.05\,mag in the $K$\ 
magnitude; and systematic errors of
about 0.11~dex, dominated by systematic effects in the temperature
scale).

We may compare these values for the surface gravities with those
deduced from the equilibrium of ionization of Fe.  On average,
abundances from Fe II lines are $0.04\pm 0.06$~dex larger than that
derived from Fe I lines. The agreement is obviously very good, but 
it could be fortuitous because other sources of errors (overionization, 
model atmospheres, $gf$, etc) may cause much larger effects 
than the mean offset measured.  The
star-to-star scatter in the residuals is quite large (0.30~dex) due to
the very limited number of Fe II lines typically used in the analysis.

Microturbulence velocities $v_t$\ were determined by eliminating
trends in the relation between expected line strength and abundances
(see Magain 1984). For stars with less than 25 lines measured we generally
adopted a value of 1.6~km s$^{-1}$ (similar to the average of the other
star), save for a few cases where obvious trends were present. Given
the typical uncertainties in the slope of expected line strength vs
abundances, this would imply an expected random error in the
microturbulence velocities of $\pm 0.15$~km s$^{-1}$. 
However, we warn the reader not to rely too much on our microturbulent values,
 because they are model dependent.

Finally, model metal abundances were set in agreement with the average 
derived Fe abundance. The adopted model atmosphere parameters are listed in Table~\ref{t:giraffeatmo}.

\begin{table*}
\begin{center}
\caption{Uncertainties in Fe abundances for stars observed with Giraffe}
\begin{tabular}{lrrrrrrrr}
\hline
Element   &Average &$T_{\rm eff}$&$\log{g}$& $[$A/H$]$ & $v_t$   &  EWs  &Total   &Total     \\
           &n. lines&         &         &         &         &       &Internal&Systematic\\
\hline
Variation  &        &    100  &  +0.30  &  +0.20  &  +0.20  &       &       &       \\
Internal   &        &     17  &   0.02  &   0.00  &   0.15  & 0.200 &       &       \\
Systematic &        &     67  &   0.11  &   0.04  &   0.04  &       &       &       \\
\hline
$[$Fe/H$]$I  & 32.0   &   0.005 &   0.070 &   0.057 &$-$0.112 & 0.035 & 0.080 & 0.037 \\
$[$Fe/H$]$II &  1.7   &$-$0.195 &   0.198 &   0.091 &$-$0.041 & 0.152 & 0.164 & 0.176 \\
$[$O/Fe$]$   &  1.8   &   0.027 &   0.052 &   0.021 &   0.103 & 0.149 & 0.172 & 0.075 \\
$[$Na/Fe$]$  &  2.0   &   0.092 &$-$0.073 &$-$0.060 &   0.040 & 0.141 & 0.150 & 0.095 \\
$[$Mg/Fe$]$  &  2.5   &$-$0.035 &$-$0.014 &$-$0.028 &   0.077 & 0.126 & 0.144 & 0.065 \\
$[$Si/Fe$]$  &  4.9   &$-$0.120 &   0.018 &$-$0.010 &   0.089 & 0.090 & 0.120 & 0.093 \\
$[$Ca/Fe$]$  &  3.9   &   0.116 &$-$0.108 &$-$0.037 &$-$0.018 & 0.101 & 0.110 & 0.100 \\
$[$Sc/Fe$]$  &  3.8   &$-$0.031 &   0.055 &   0.014 &   0.048 & 0.103 & 0.115 & 0.057 \\
$[$Ti/Fe$]$  &  1.0   &   0.170 &$-$0.058 &$-$0.020 &$-$0.088 & 0.200 & 0.216 & 0.148 \\
$[$V/Fe$]$   & 10.3   &   0.195 &$-$0.070 &$-$0.007 &$-$0.100 & 0.062 & 0.109 & 0.138 \\
$[$Cr/Fe$]$  &  2.9   &   0.145 &$-$0.045 &$-$0.020 &$-$0.076 & 0.117 & 0.138 & 0.113 \\
$[$Ni/Fe$]$  & 13.2   &$-$0.043 &   0.025 &$-$0.002 &   0.040 & 0.055 & 0.072 & 0.043 \\
$[$Ba/Fe$]$  &  1.0   &   0.014 &   0.005 &   0.037 &$-$0.057 & 0.200 & 0.208 & 0.092 \\
\hline
\end{tabular}
\label{t:errorparam}
\end{center}
\end{table*}

\subsection{Fe Abundances}


Individual [Fe/H] values are listed in
Table~\ref{t:giraffeatmo}. Reference solar abundances are as in
Gratton et al. (2003b). Throughout our analysis, we use the same line
parameters discussed in Gratton et al. (2003b); in particular,
collisional damping was considered using updated constants from
Barklem et al. (2000).  

Table~\ref{t:errorparam} lists the impact of
various uncertainties on the derived abundances for the elements
considered in our analysis. Variations in parameters of the model
atmospheres were obtained by changing each of the parameters at a time
for star \#7006354, assumed to be representative of all the stars
considered in this paper. 

The first three rows of the table give the
variation of the parameter used to estimate sensitivities, the
internal (star-to-star) errors, and the systematic errors (common to
all stars) in each parameter. The second column gives the average
number of lines $n$\ used for each element. Columns 3-6 give the
sensitivities of the abundance ratios to variations of each
parameter. Column 7 gives the contribution to the error given by
uncertainties in the EWs for individual lines: this is
$0.200/\sqrt{n}$, where 0.200 is the error in the abundance derived
from an individual line, as obtained by the median error over all the
stars. The last two columns give the total internal and systematic
errors, obtained by summing quadratically the contribution of the
individual source of errors, weighted according to the errors
appropriate for each parameter. For the systematic errors, the
contribution due to equivalent widths and to microturbulence
velocities (quantities derived from our own analysis), were divided
by the square root of the number of cluster members observed. Note
that this error analysis does not include the effects of covariances
in the various error sources, which are however expected to be quite
small for the program stars.

Errors in Fe abundances from neutral lines are dominated by
uncertainties in the microturbulent velocity. We estimate random
errors of 0.091 dex, and systematic errors of 0.037 dex. From
Table~\ref{t:giraffeatmo}, the average Fe abundance from all stars of
\object{NGC 6441} is [Fe/H]=$-0.34\pm 0.02$\ (error of the mean), with an
r.m.s. scatter of 0.08 dex from 25 stars. Then, the first result of
our analysis is that the metallicity of \object{NGC 6441} is [Fe/H]=$-0.34\pm
0.02\pm 0.04$. This value agrees well, within the errors, with the
average Fe abundance determined in Paper II ([Fe/H]=$-0.39\pm 0.04\pm
0.05$). The small difference could be explained by the trends in EWs
discussed in Section 4.1. More weight should be attributed to the UVES
results because the are based on spectra of higher resolution.
Other literature determinations of the metal abundance of
\object{NGC 6441} are discussed in Paper II: they generally agree on a high
metal abundance for this cluster.

It should be noted that the observed star-to-star scatter in Fe abundance 
is actually
smaller than the estimate of the random errors. Hence, present data,
from a wider sample than considered in Paper II, do not support the
existence of a metal abundance spread in \object{NGC 6441}. Notably, there is no
possible member of \object{NGC 6441} that is significantly more metal-poor than
the average abundance for the cluster. About 12\% of the horizontal
branch stars in \object{NGC 6441} are on the blue side of the RR Lyrae
instability strip, and an additional 4\% are close to or within the
instability strip (see Section 5 for a description of how we estimated
these fractions): it could be claimed that these stars belong to a
more metal-poor population. Clementini et al. (2005) have shown that
the mean metallicity of the RR Lyrae of \object{NGC 6441} is high, close to the
mean value for the cluster. We may add here that the lack of any star
more metal-poor than [Fe/H]=$-0.6$\ in our sample of bona fide members
of \object{NGC 6441} (30 stars if we include also 
the stars observed with UVES)
excludes at the 99\% level of confidence the existence of a
population of metal-poor stars as large as 16\% of the cluster
population (the total number of HB stars bluer than red HB).
We may also exclude at 97\% level of confidence a metal poor population 
accounting for 12\% of the stars, that is the fraction of blue HB stars over
the total number of HB stars.  
While a small population of metal-poor stars, enough to
justify the population of blue horizontal branch stars, may still be
present in the cluster, it is clear that at least the cluster RR Lyrae
must result from the evolution of stars with the typical 
abundance found for \object{NGC 6441}. We conclude that very likely the
anomalous HB and the spread in color of the RGB of \object{NGC 6441} are not
related to a spread of abundance for the heavy elements. While the
first is probably a manifestation of the second parameter problem, the
second is likely due only to differential reddening and extensive
contamination by field (bulge) stars.

\begin{table*}
\begin{center}
\caption{Average abundances for \object{NGC 6441}.}
\scriptsize
\begin{tabular}{lrrrrrrrrrrrrrr}
\hline
Star ID &
\multicolumn{3}{c}{$[$Mg/Fe$]$}&
\multicolumn{3}{c}{$[$Si/Fe$]$}&
\multicolumn{3}{c}{$[$Ca/Fe$]$}&
\multicolumn{3}{c}{$[$Sc/Fe$]$}&
\multicolumn{2}{c}{$[$Ti/Fe$]$}\\
\hline
6005198&2&0.28&0.08&7&   0.56&0.46&4&$-$0.02&0.16&5&   0.05&0.18&1&   0.48\\      
6005934&3&0.26&0.22&7&   0.50&0.33&4&   0.31&0.31&5&   0.10&0.14&1&   0.12\\ 
6007741&2&0.44&0.33&1&   0.60&    &3&   0.40&0.06&2&   0.38&0.22& &       \\ 
6010149&1&0.35&    &5&   0.53&0.32& &       &    &2&       &    & &       \\ 
6012768&2&0.13&0.10&2&   0.32&0.03&4&   0.13&0.19&2&$-$0.22&0.09&1&   0.28 \\
7004955&2&0.38&0.09&2&   0.68&0.28&4&   0.43&0.18&2&   0.25&0.17&1&   0.60 \\
7006255&3&0.14&0.36&7&   0.22&0.18&3&   0.18&0.18&5&   0.12&0.18&1&   0.17 \\
7006305&3&0.54&0.03&7&   0.38&0.27&4&   0.35&0.22&5&$-$0.14&0.18&1&   0.39 \\
7006319&3&0.50&0.18&7&   0.45&0.18&4&   0.24&0.19&5&   0.28&0.12&1&   0.38 \\
7006354&2&0.32&0.08&2&   0.57&0.12&4&   0.05&0.28&2&$-$0.21&0.18&1&   0.43 \\
7006377&3&0.20&0.22&7&   0.33&0.32&4&   0.10&0.15&5&$-$0.02&0.16&1&   0.07 \\
7006470&3&0.35&0.27&7&   0.24&0.22&3&   0.17&0.15&4&$-$0.03&0.49&1&   0.35 \\
7006590&3&0.27&0.41&6&   0.19&0.21&4&$-$0.13&0.22&5&   0.10&0.22& &        \\
7006935&3&0.30&0.17&7&   0.19&0.18&4&$-$0.03&0.16&5&$-$0.02&0.21&1&   0.09\\
7006983&3&0.57&0.05&7&   0.43&0.20&4&   0.43&0.15&5&   0.21&0.17&1&   0.55 \\
7007064&3&0.31&0.09&7&   0.35&0.21&4&$-$0.16&0.21&5&$-$0.18&0.15&1&   0.18\\
7007118&3&0.40&0.20&7&   0.33&0.19&4&   0.30&0.22&5&   0.17&0.21&1&   0.53 \\
7007884&2&0.53&0.09&1&   0.40&    &4&$-$0.05&0.22&2&   0.07&0.27&1&   0.05 \\
7008891&2&0.33&0.07&2&$-$0.09&0.11&4&   0.32&0.12&2&   0.34&0.00&1&   0.47 \\
7013582&2&0.60&0.02&2&   0.48&0.07&4&   0.48&0.21&2&   0.09&0.25&1&$-$0.03 \\
8005158&2&0.49&0.09&2&   0.65&0.07&4&   0.15&0.38&2&$-$0.03&0.19&1&   0.21 \\
8005404&3&0.58&0.05&7&   0.55&0.26&4&   0.19&0.17&4&   0.20&0.38&1&   0.23 \\
8006535&3&0.57&0.37&7&   0.55&0.19&4&   0.22&0.28&5&   0.30&0.06&1&   0.41 \\
8008693& &    &    &4&   0.20&0.50&4&   0.45&0.27&5&$-$0.02&0.10&1&   0.58 \\
8013657&2&0.31&0.06&2&   0.68&0.03&4&   0.37&0.19&2&$-$0.09&0.35&1&   0.62 \\
\hline
Star ID&
\multicolumn{3}{c}{$[$V/Fe$]$}&
\multicolumn{3}{c}{$[$Cr/Fe$]$}&
\multicolumn{3}{c}{$[$Ni/Fe$]$}&
\multicolumn{2}{c}{$[$Ba/Fe$]$}\\  
\hline
6005198&10&$-$0.86&0.31&4&$-$0.32&0.49&13&   0.15&0.22&1& 0.20\\  
6005934& 9&$-$0.17&0.32&4&   0.14&0.43&15&   0.07&0.19&1& 0.16\\
6007741&  &       &    &1&   0.02&    &11&   0.25&0.25&1& 1.12\\
6010149&10&   0.51&0.40&3&$-$0.05&0.20& 4&   0.39&0.26& &      \\    
6012768&  &       &    & &       &    &12&   0.16&0.15&1& 0.23\\
7004955&  &       &    &1&$-$0.13&    &12&   0.25&0.25&1& 0.31\\
7006255&10&$-$0.19&0.27&4&$-$0.17&0.24&15&   0.05&0.26&1& 0.04\\
7006305&11&$-$0.06&0.45&4&$-$0.20&0.37&15&   0.36&0.27&1& 0.11\\
7006319&11&   0.35&0.28&4&   0.08&0.23&16&   0.02&0.27&1& 0.09\\
7006354&  &       &    &1&   0.20&    &11&   0.14&0.23&1& 0.19\\
7006377&10&$-$0.21&0.15&3&$-$0.46&0.17&16&   0.10&0.20&1& 0.01\\
7006470& 6&$-$0.26&0.21&4&$-$0.23&0.11&15&   0.27&0.28&1& 0.13\\
7006590&11&$-$0.10&0.22&4&$-$0.20&0.29&15&$-$0.13&0.16&1&$-$0.24\\
7006935&11&$-$0.19&0.25&4&$-$0.29&0.13&16&$-$0.03&0.19&1&$-$0.18\\
7006983&11&   0.08&0.22&4&$-$0.25&0.23&13&   0.25&0.17&1& 0.17\\
7007064&11&$-$0.14&0.25&4&$-$0.18&0.24&16&   0.12&0.21&1&$-$0.01\\
7007118&11&   0.06&0.32&4&$-$0.15&0.25&16&   0.14&0.22&1& 0.06\\
7007884&  &       &    &1&$-$0.13&    &11&$-$0.04&0.27&1& 0.21\\
7008891&  &       &    &1&   0.10&    &12&   0.02&0.21&1& 0.23\\
7013582&  &       &    &1&   0.09&    &11&   0.24&0.16&1&$-$0.08\\
8005158&  &       &    &1&$-$0.30&    &12&   0.11&0.12&1& 0.02\\
8005404&11&   0.29&0.33&4&$-$0.17&0.19&14&$-$0.10&0.14&1& 0.08\\
8006535&11&   0.16&0.29&4&$-$0.20&0.43&15&   0.17&0.20&1& 0.14\\
8008693&10&$-$0.71&0.37&4&   0.15&0.37&13&   0.56&0.36&1& 0.08\\
8013657&  &       &    &1&   0.11&    &10&   0.22&0.26&1& 0.39\\
\hline
\end{tabular}
\normalsize
\label{t:abundgc6441}
\end{center}
\end{table*}

\begin{table}
\begin{center}
\caption{Abundances for \object{NGC 6441} members}
\normalsize
\begin{tabular}{lccccc}
\hline
El. &\multicolumn{3}{c}{This Paper}&\multicolumn{2}{c}{Paper II}\\
        & n.stars & [A/Fe] & r.m.s. & [A/Fe] & r.m.s. \\
\hline
Mg & 24 &  +0.38 & 0.14 & +0.34 & 0.09 \\                      
Si & 25 &  +0.41 & 0.19 & +0.33 & 0.11 \\
Ca & 24 &  +0.21 & 0.19 & +0.03 & 0.04 \\
Sc & 24 &  +0.07 & 0.17 & +0.15 & 0.15 \\
Ti & 22 &  +0.33 & 0.20 & +0.29 & 0.10 \\
V  & 16 &  +0.01 & 0.24 & +0.29 & 0.14 \\ 
Cr & 24 &$-$0.11 & 0.18 & +0.15 & 0.18 \\
Ni & 24 &  +0.13 & 0.13 & +0.13 & 0.07 \\ 
Ba & 23 &  +0.10 & 0.14 & +0.17 & 0.13 \\ 
\hline
\end{tabular}
\label{t:meanabund}
\end{center}	
\end{table}

\subsection{Abundances for other elements} 


Table~\ref{t:abundgc6441} lists the average cluster abundances for the individual
elements. For each star and for each element, we give the number of
lines used in the analysis, the average abundance, and the
r.m.s. scatter of individual values. Abundances for the odd elements
of the Fe group (Sc and V) were derived with consideration for the not
negligible hyperfine structure of these lines (see Gratton et
al. 2003b for more details).

Average abundances for the cluster, as well as the r.m.s. scatter of
individual values around this mean value, are given in
Table~\ref{t:meanabund}. For comparison, we also give the values
derived from analysis of the UVES spectra in Paper II. In general, the
values for the scatter agree fairly well with those estimated in our
error analysis.

The overall pattern of abundance of \object{NGC 6441} is typical of a
globular cluster, with a large excess of the $\alpha-$elements. This
excess is larger than usually found in stars of similar metallicity
belonging to the thick disk, where values of [Mg/Fe] and
[Si/Fe] of about $\sim 0.2$~dex are generally found (Gratton et
al. 2003c; Bensby et al. 2005; Soubiran \& Girard 2005). This large excess
of the $\alpha-$elements
suggests that the material from which the stars of \object{NGC 6441} formed was
enriched by massive core collapse SNe, with little if any contribution
by type Ia SNe. This might suggests either a peculiar metal enrichment
process, or a very old age for the cluster. Unfortunately, no accurate age
derivation is yet available and such estimate would be
anyway difficult to be derived, due to the impact of differential
reddening.

\section{THE O-NA ANTICORRELATION}


The Na abundances were derived from the 6154-60~\AA\ doublet alone;
they include corrections for departures from LTE, following the
treatment by Gratton et al. (1999). We prefer not to use the
5682-88~\AA\ doublet, that is very strong in the spectra of the
program stars, because these lines are heavily saturated, contaminated
by blends at the resolution of Giraffe spectra, and affected by large
deviations from LTE. Additionally, the S/N of the spectra obtained
with the HR11 grating is much lower than those obtained with HR13
grating, so that addition of the 5682-88~\AA\
doublet data does not improve our Na abundances.

Telluric absorption lines were removed from the spectra in the region
around the [O{\sc I}] lines. No attempt was made to remove the strong
auroral emission line; due to the combination of the Earth and stellar
motions at the epoch of observations, the auroral emission line
typically is at a wavelength about 0.7~\AA\ blueward of the stellar
line, so that it only occasionally disturbs its profile 
at the resolution of the Giraffe spectra. The O abundances were
derived from equivalent widths: and confirmed by
spectral synthesis. We did not apply any correction for either the
blending with the Ni~I line at 6300.339~\AA, or for formation of CO. The
Ni{\sc I} line is expected to contribute about 4~m\AA\ to the EW
of the [O{\sc I}] line, using the line parameters by Allende Prieto et
al. (2001); the corresponding correction to O abundances is about 0.05
dex downward. CO coupling should be strong at the low effective
temperature of the program stars. Unfortunately, the abundance of C is not
determined. However, we expect that C is strongly depleted in stars
near the tip of the RGB, with expected values of
[C/Fe]$\sim -0.6$\ (Gratton et al. 2000). If the C abundance follows the Fe
one in unevolved stars (as usually observed in metal-rich environments: see e.g. Gratton
et al. 2000), we expect typical values of [C/O]$\sim -0.8$\ for
stars in \object{NGC 6441}, in agreement with the non-detection of the C$_2$\ lines
in the spectral range 5610-30~\AA. Neglecting CO formation, we could have
underestimated the O abundances from forbidden lines by $\sim
0.05$~dex. These two corrections should then roughly compensate,
 anyway they are within the error bars of the present
determinations.

\begin{figure}[h]
\includegraphics[width=8.8cm]{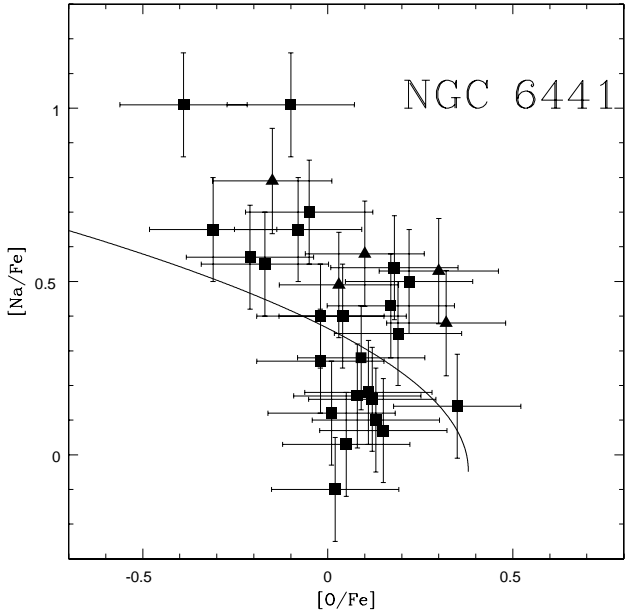}
\caption[]{[Na/Fe] vs [O/Fe] abundance ratios for the stars 
adopted as member of the cluster. Triangles 
are objects observed with FLAMES-UVES (see Paper II),
while squares are those of the present sample.
The line represents the mean loci of the Na-O anticorrelation as derived from data of several GCs.}
\label{f:anti}
\end{figure}

\begin{figure}[h]
\includegraphics[width=8.8cm]{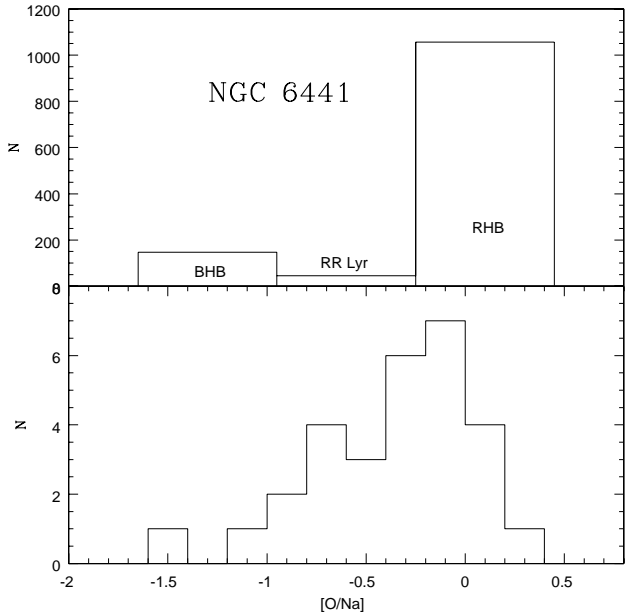}
\caption[]{Upper panel indicates the incidence of three different stars populations
along the HB, selected from
the CMD. Lower panel shows the distribution of the [O/Na] abundance ratios.}
\label{f:distnao}
\end{figure}

Abundances of O and Na are listed in Table~\ref{t:giraffeatmo}. The
[Na/Fe] ratio as a function of [O/Fe] ratio is displayed in Figure \ref{f:anti},
where we also plotted (with different symbols) the results obtained
from the UVES spectra in Paper II. Overplotted is the mean line for several GCs (see Carretta et
al. 2006a). The usual Na-O anticorrelation seen in several other GCs
is also evident in \object{NGC 6441}. Most of the red giants of \object{NGC 6441} have
rather high O abundances, but there exists a substantial fraction of
O-poor, Na-rich stars. The Na abundances observed in the O-poor stars
of \object{NGC 6441} are very large, possibly larger than in other GCs. The
distribution function of stars along the Na-O anticorrelation in
NGC~6441 is shown in the lower panel of 
 Figure \ref{f:distnao}, where the ratio [O/Na] from our data is
used. The histogram shows a a peak at rather large [O/Na] values,
including about 2/3 of the stars, with an extended tail down to very
low [O/Na] values, including the remaining one third of the stars.

\begin{figure}[h]
\includegraphics[width=8.8cm]{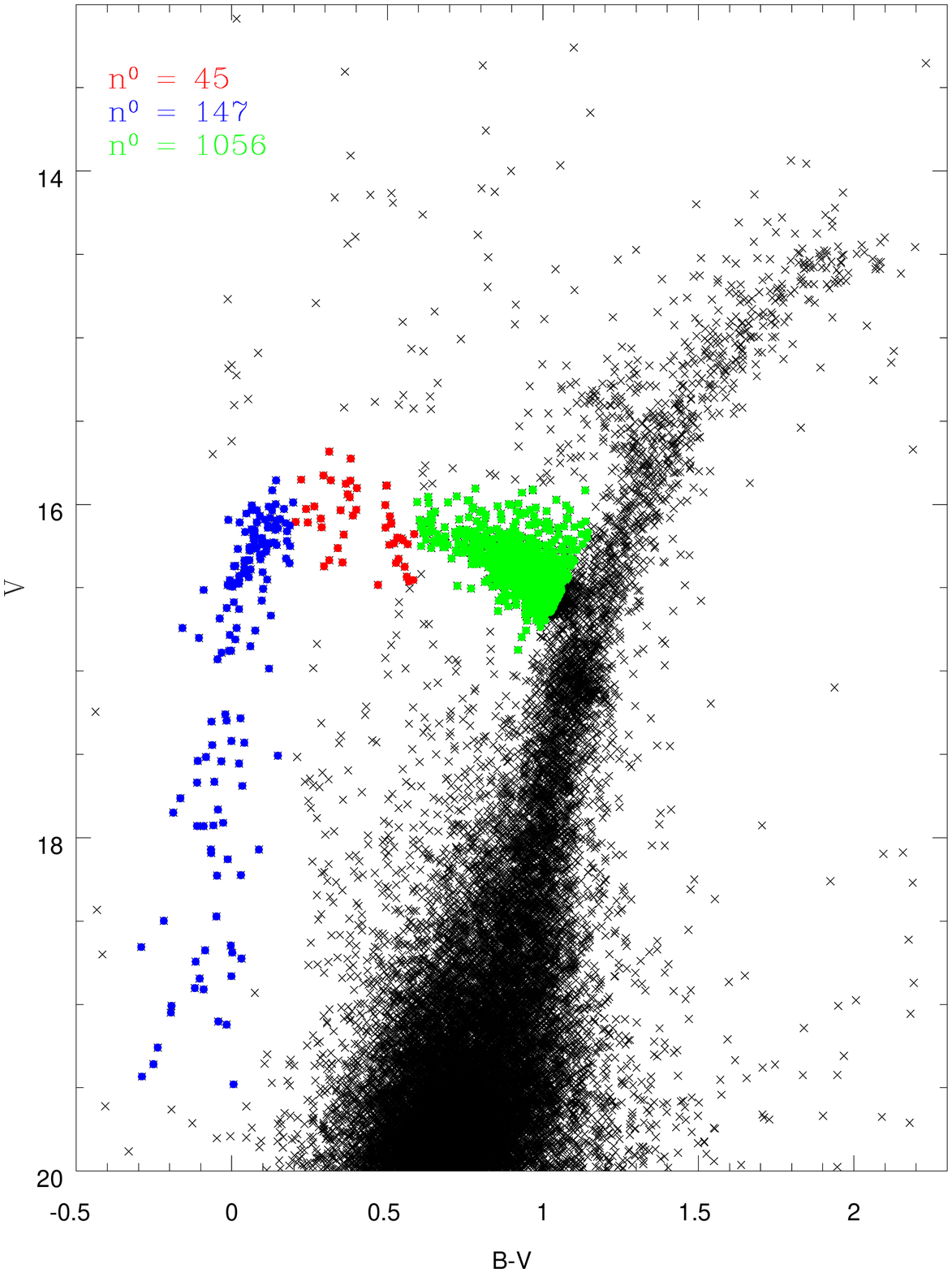}
\caption[]{Dereddened CMD diagram from HST data of the central region of \object{NGC6441} (Piotto et al. 2002). 
The different colors indicate the areas on the CMD in which the populations selected for the comparison can be found. 
From top to bottom: RR Lyr, blue HB, 
and red HB.}
\label{f:cmdbox}
\end{figure}

We may compare this distribution of the [O/Na] values with the colors
of the stars along the HB of \object{NGC 6441} as shown in the HST CMD 
of the inner region of the cluster (Rich et al. 1997). 
We counted the cluster stars populating the three different parts of the HB, 
i.e. the blue HB, the RR Lyr instability strip and the red HB. The areas of the CMD selected 
to represent each of the HB part are shown in Figure \ref{f:cmdbox}.  
The final comparison is shown in the top panel of Figure \ref{f:distnao}.

 While this comparison is
essentially qualitative (stars along the HB are binned in broad bins
of colors, and both distributions have to be transformed into a mass
distribution for the comparison to be really meaningful), the two
distributions appear to be quite similar. About 3/4 of the RGB stars
of \object{NGC 6441} (21 out of 29) are O-rich and Na-poor, a fraction
similar to the RHB over the total of HB stars (1056 out of 1248). Most
of the remaining RGB stars have intermediate composition, with only
one example of very low [O/Na] ratios. This should correspond to the
 population of HB stars of intermediate colors (possibly falling
within the RR Lyrae instability strip), and to the tail
including $\sim$15\% of the stars on the blue part of the
HB. Summarizing, the present data supports a qualitative agreement between
the distribution of [O/Na] ratios for RGB stars and of colors along
the HB of \object{NGC 6441}. This is in agreement with a scenario where the
distribution of [O/Na] ratios in a GC reflects a distribution of He
contents, and of masses of both RGB and HB stars.

\begin{table}
\begin{center}
\caption{EWs for neutron capture element lines in the Ba-star \#6007741 in \object{NGC 6441}}
\begin{tabular}{lccccc}
\hline
Element & Wavelength & E.P. & $\log{gf}$ &ref. & EW \\
      & (\AA)   & (eV) &       &               &(m\AA) \\
\hline
Y~I   & 6222.58 & 0.00 & -1.70 & 1 &   78.0 \\
Y~II  & 5728.89 & 1.84 & -1.12 & 1 &   50.0 \\
Zr~I  & 6127.46 & 0.15 & -1.06 & 2 &  145.0 \\
Zr~I  & 6134.57 & 0.00 & -1.28 & 2 &  134.0 \\
Zr~I  & 6140.46 & 0.52 & -1.41 & 2 &  100.0 \\
Zr~I  & 6313.03 & 1.58 &  0.27 & 2 &   86.0 \\
Ba~II & 6141.75 & 0.70 &  0.00 & 3 &  382.1 \\
La~II & 5769.06 & 1.25 & -0.69 & 4 &   88.0 \\
La~II & 5797.57 & 0.24 & -1.36 & 4 &  107.0 \\
La~II & 6390.48 & 0.32 & -1.41 & 4 &   90.0 \\
Ce~II & 5613.69 & 1.42 & -0.47 & 5 &   35.0 \\
Nd~II & 5740.86 & 1.16 & -0.53 & 6 &   54.0 \\
Nd~II & 5804.00 & 0.74 & -0.53 & 6 &   69.0 \\
Nd~II & 5811.57 & 0.86 & -0.86 & 6 &   50.0 \\
Nd~II & 5825.86 & 1.08 & -0.66 & 6 &   52.0 \\
Nd~II & 6382.06 & 1.44 & -0.75 & 6 &   46.0 \\
Eu~II & 6313.03 & 1.28 & -1.02 & 7 &$<$20.0 \\
\hline
\end{tabular}
\label{t:ncaptureew}\\
1. Hannaford et al. 1982
2. Bi\'emont et al. 1981
3. Holweger \& M\"uller 1974
4. Lawler et al. 2001b
5. Bi\'emont et al. 2005
6. den Hartog et al. 2003
7. Lawler et al. 2001a
\end{center}
\end{table}

\begin{table}
\begin{center}
\caption{Abundances for neutron capture elements in the Ba-star \#6007741 in \object{NGC 6441}}
\begin{tabular}{lcccc}
\hline
Element & n. Lines & $[$A/Fe$]$ & r.m.s. & solar fraction   \\
        &          &        &        & (main component \\
        &          &        &        & of the s-process) \\
\hline
$[$Y/Fe$]$~I   & 1 &   $+$0.84 &      & 0.81 \\
$[$Y/Fe$]$~II  & 1 &   $+$0.64 &      & 0.81 \\
$[$Zr/Fe$]$~I  & 4 &   $+$0.57 & 0.18 & 0.66 \\
$[$Ba/Fe$]$~II & 1 &   $+$1.12 &      & 0.89 \\
$[$La/Fe$]$~II & 3 &   $+$1.19 & 0.32 & 0.75 \\
$[$Ce/Fe$]$~II & 1 &   $+$0.93 &      & 0.78 \\
$[$Nd/Fe$]$~II & 5 &   $+$0.73 & 0.28 & 0.45 \\
$[$Eu/Fe$]$~II & 1 &  $<+$0.93 &      & 0.07 \\
\hline
\end{tabular}
\label{t:ncaptureabu}
\end{center}	
\end{table}

\section{THE BA-STAR \#6007741}


\begin{figure}[h]
\includegraphics[width=8.8cm]{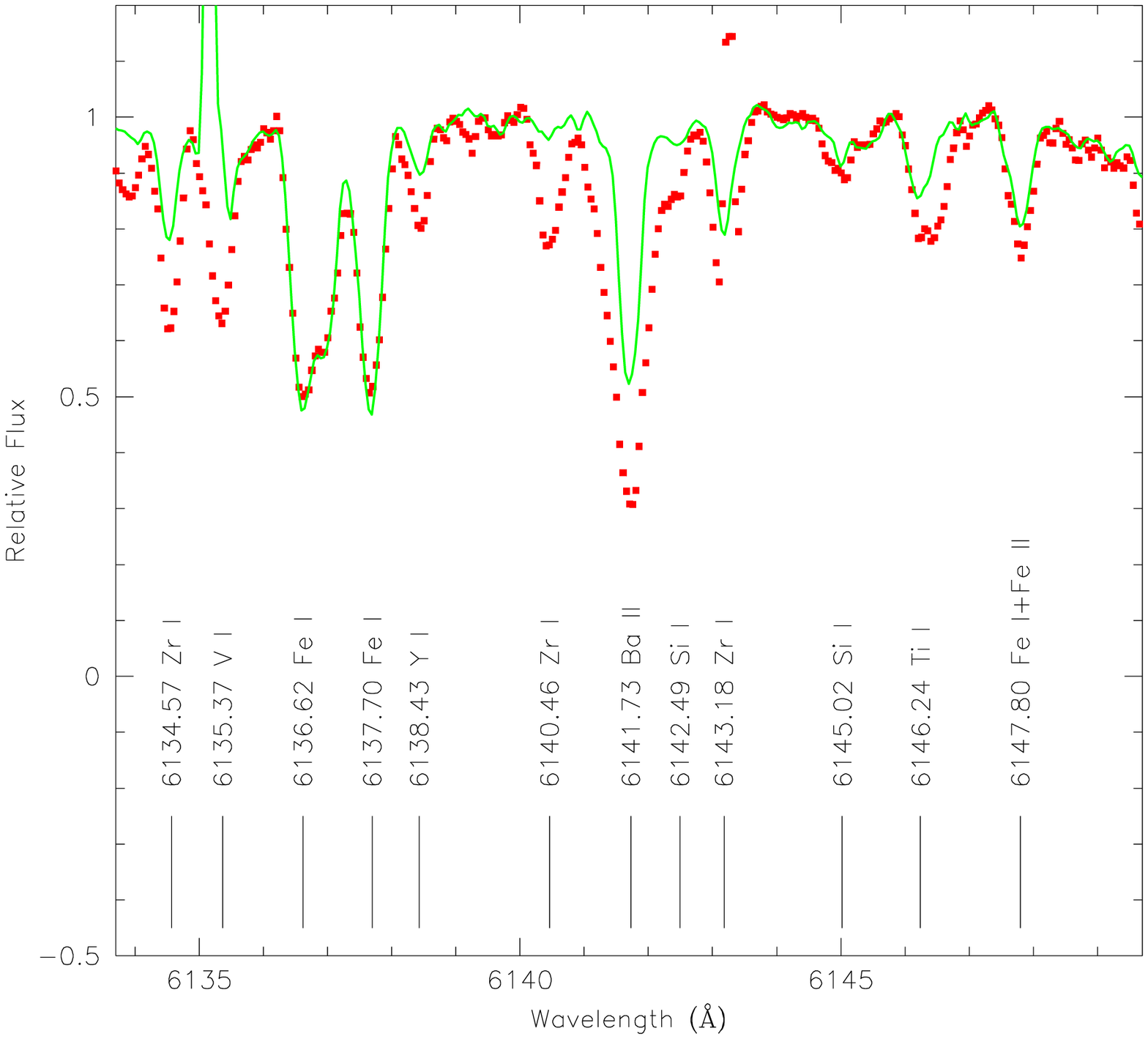}
\caption[]{Comparison between the spectrum of stars \#6004471 (dots) and
\#7006935 (solid line) in the region of the Ba~II line at 6141.73~\AA.
The two stars have similar atmospheric parameters.
The first is characterized by much stronger lines of the n-capture elements.}
\label{f:bastar1}
\end{figure}

\begin{figure}[h]
\includegraphics[width=8.8cm]{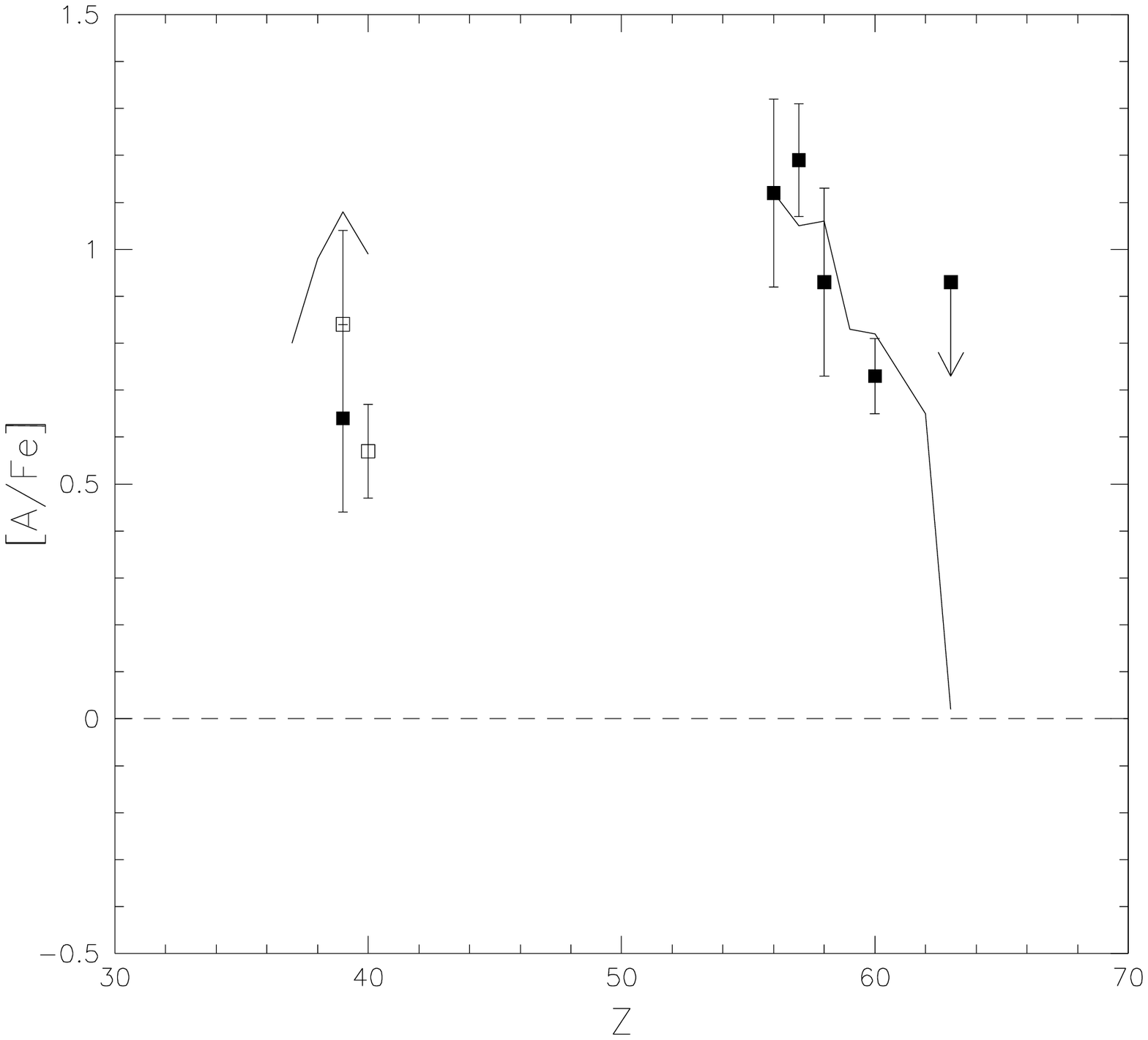}
\caption[]{n-capture elements abundances with respect to iron measured in star \#6007741. The line indicates the 
scaled s-process solar pattern. The Eu abundance shown is an upper limit. Open
symbols are abundances from neutral lines; filled symbols from singly ionized
lines.}
\label{f:sel}
\end{figure}

One member of \object{NGC 6441}, star \#6007741, turned out to have a
very strong Ba line (see Figure \ref{f:bastar1}). We measured EWs for a few
additional lines of neutron capture element on the spectrum of this
star; they are listed in
Table~\ref{t:ncaptureew}. Table~\ref{t:ncaptureabu} lists the average
abundances for the n-capture elements obtained from these EWs, along
with the s-fraction (due to the main component of the s-process) in
the Sun according to K\"appeler et al. (1989, 1990a, 1990b). The large
overabundances of Ba and La (mainly s-process elements) and the much
lower overabundances of the mainly r-process elements Nd and Eu, which are
only slightly larger than those found in the remaining stars of \object{NGC 6441}
(see Paper II), identify this as an S-star. Figure \ref{f:sel} shows how well the
abundance pattern of this star reproduces the abundances attributed to
the main component of the s-process in the Sun for what concern the
Ba-peak; however, we notice that the overabundances of Y and Zr are
about two times smaller than the value expected from this distribution.

This abundance pattern strongly suggests that star \#6007741 has been
enriched by material processed during the thermal pulses phase of a
small mass AGB star. The star itself is too faint for having
experienced thermal pulses: the metallicity of \object{NGC 6441} is similar to
that of the LMC, and we know that in the LMC thermal pulses occur
during the evolution of stars in the approximate mass range
$1.2<M<3~M_\odot$, and with a luminosity of $-3.5>M_{Bol}>-6$
(Frogel et al. 1990). Hence,
the abundance pattern observed in \#6007741 is due to mass transfer
from an originally more massive companion. Note that the system may
have been disrupted after the mass transfer episode by close
encounters with other cluster members, so it is not obvious that
\#6007741 should still have a companion. However, although the time span 
covered by our observations is quite limited, we detect a 
small variation in radial velocity for this object.

\begin{figure*}[h]
\includegraphics[width=18cm]{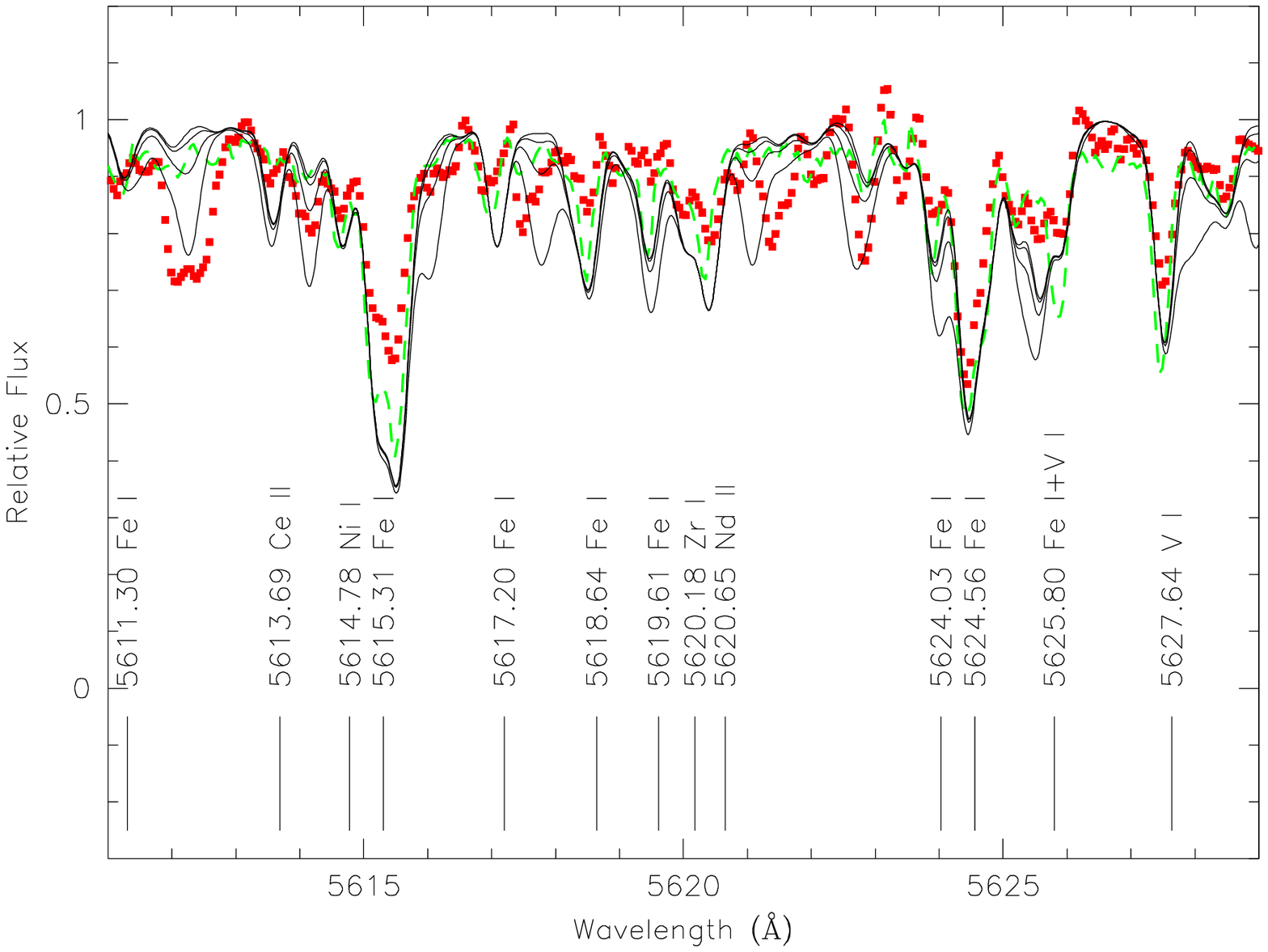}
\caption[]{Synthesis of the C$_2$ Swan band in the 5610-5630 \AA\ wavelength range for
star \#6007741 (dots). The spectrum of star \#7006935 is also plotted for
comparison (dashed line). Synthetic spectra (solid lines) were computed for
atmospheric parameters appropriate to star \#6007741, N and O abundances of
[N/Fe]=0.5 and [O/Fe]=0, and C abundances of [C/Fe]=$-$0.2, 0, 0.2, and 0.3 dex
respectively. }
\label{f:c2}
\end{figure*}

The donor star is expected to have produced also large amounts of C during
He-shell flashes. We have derived the C abundance for star
\#6007741 using the spectral region 5610-5630~\AA, which includes
several lines of the 0-1 vibrational band of the C$_2$\ Swan
system. We compiled a line list for this spectral region from Kurucz
(1992); for C$_2$, we used the line list from Phillips \& Davis
(1968), the electronic oscillator strength from Lambert (1978), the
Franck-Condon factor from Dwivedi et al. (1978), and the H\"onl-London
factors from Schadee (1964). We verified that a synthetic spectrum
computed with this line list gives a very good fit to the solar
spectrum for the standard solar C abundance. We then computed
synthetic spectra appropriate for \#6007741, and compared this
spectrum with observations. In spite of the rather low S/N of the
spectra, C$_2$\ lines were clearly detected in the spectrum of star
\#6007741 (see Figure \ref{f:c2}); the best fit is obtained with a C/O ratio of
$0.87\pm 0.10$\ (there is a strong coupling between C and O abundances
through formation of CO at the temperature of this star). Since it is
expected that C is strongly depleted in bright RGB stars of GCs (and
this indeed occurs for the other program stars, where no C$_2$\ lines
are detectable: see for instance the case of star \#7006935 shown in the
same Figure \ref{f:c2}), we conclude that \#6007741
has an anomalously high C abundance. While we have not observed the
G-band of this star, we might predict that \#6007741 should have a
rather strong G-band, and should then be classified as a mild CH-star.

Ba and CH-stars are rare in globular clusters (for a catalogue, see
Bartkevicius 1996). Most of them are in GCs of low central
concentration. The rarity of Ba and CH-stars in GCs might be related
with the high probability that primordial binaries with the right
separation to form Ba and CH-stars (McClure et al. 1980; McClure 1984)
are destroyed in the dense environments of GCs. It was quite
unexpected to find such a star in \object{NGC 6441}, that is a quite strongly
concentrated GC ($c=1.85$: Harris 1996).

\section{CONCLUSIONS}


In this paper we have derived RVs, atmospheric
parameters and elemental abundances for a number of red giants in the
field of the globular cluster \object{NGC 6441}, observed with the FLAMES
multifibre facility and the Giraffe spectrograph at VLT2. Membership
of the stars was derived from location in the cluster, radial
velocities, and taking into account their chemical composition: the
final sample includes 25 stars that are likely members of the cluster
(in addition to the 5 stars observed with the UVES spectrograph and
discussed in paper I). Atmospheric parameters were obtained from the
photometry: temperatures were obtained from the 2MASS $K$ magnitudes,
exploiting an average $K-(V-K)$\ relation valid for cluster stars. In
this way we minimize the impact of differential reddening on abundances.

From the analysis of the Giraffe spectra we derived an average
metallicity of [Fe/H]=$-0.34\pm 0.02\pm 0.04$~dex, slightly higher
than the value obtained in Paper II from UVES data. There is no
indication of star-to-star scatter larger than the observational
errors. The possibility that the RR Lyrae and the blue HB stars (unexpected in such a metal rich cluster) 
are due to a population of metal poor objects can then be ruled out at a high level of 
confidence.
 The cluster is overabundant in the $\alpha-$elements Mg, Si,
Ca, and Ti, indicating enrichment by massive core collapse SNe.

We measured O and Na abundances for 24 stars from the forbidden [O{\sc I}]
lines at 6300.3, 6363.8~\AA\ and the Na doublet at
6154-60~\AA. Combining this data with those extracted from UVES
spectra (Paper II), O and Na abundances are available for 29 stars. The
[Na/Fe] versus [O/Fe] ratios follow the well known Na-O
anticorrelation, signature of proton-capture reactions at high
temperatures, found in all other GCs examined so far. The
distribution function of stars in [O/Na] (i.e.along the Na-O
anticorrelation) is dominated by O-rich, Na poor stars, which
constitute about 70\% of the cluster stars, with a tail toward lower O
and higher Na abundances. This distribution appear to be similar
to the color distribution of stars along the HB, dominated by a well
populated RHB, but with a significant tail of bluer stars. Adequate
modelling is required to show whether qualitative agreement
corresponds indeed to a quantitative one. This on turn should likely require
an age determination for \object{NGC 6441}.
 
One of the star (\#6007741) turned out to be a Ba-star, a class of
objects that is relatively rare in globular clusters. The distribution
of elements around the Ba-peak reproduces well the abundance pattern
expected for the main component of the s-process; the lighter elements
Y and Zr are however less overabundant by about a factor of
two. Star \#6007741 also has a rather high C abundance (the C/O factor is
close to 1), in agreement with a scenario where the overabundance of
s-process elements is due to mass transfer from a thermally pulsing
small mass AGB star.

\begin{acknowledgements}
This research has been funded by grant "Experimental Nucleosynthesis
in Clean Environments" by INAF. This publication makes
use of data products from the Two Micron All Sky Survey, which is a
joint project of the University of Massachusetts and the Infrared
Processing and Analysis Center/California Institute of Technology,
funded by the National Aeronautics and Space Administration and the
National Science Foundation
\end{acknowledgements}

\end{document}